\documentclass[preprint,showpacs,preprintnumbers,aps,nofootinbib]{revtex4}
\usepackage{amssymb}
\usepackage{graphicx}

\begin{document}

\title{Nonforward Compton scattering in AdS/CFT correspondence}
\author{Jian-Hua Gao}
\email{gaojh79@ustc.edu.cn}
\affiliation{Department of Modern Physics, University of Science and Technology of China,
Hefei, Anhui 230026, People's Republic of China}
\author{Bo-Wen Xiao}
\email{bxiao@lbl.gov}
\affiliation{Nuclear Science Division, Lawrence Berkeley National Laboratory, Berkeley,
California 94720, USA}
\date{\today}

\begin{abstract}
We study the nonforward Compton scattering, in particular, the deeply virtual
Compton scattering from AdS/CFT. We first calculate the contributions from
the \textit{s}-channel and \textit{u}-channel supergravity diagrams as well as the four-point interaction diagram
 which correspond to the Compton scatterings on a dilaton target in CFT.
 Furthermore, we study the Compton scattering on a dilatino target. Assuming that protons can be
identified as supergravity modes of the dilatino, we compare the calculated
deeply virtual Compton scattering
cross section to the low-energy experimental data from the H1 and ZEUS
collaborations and find good agreement. We also discuss the \textit{t}-channel
graviton exchange contribution and show that it should be dominant in the
high-energy limit.
\end{abstract}

\maketitle

\section{Introduction}

A decade ago, the AdS/CFT correspondence\cite%
{Maldacena:1997re,Witten:1998qj,Gubser:1998bc} was conjectured.
Since then, it has provided us with new insights into gauge theories in
a strong coupling regime. There has been substantial progress in
studying strong coupling gauge theories by applying this
technique. In particular, Polchinski and Strassler
\cite{Polchinski:2001tt, Polchinski:2002jw} studied the deep
inelastic scattering (DIS) of hadrons by using the correspondence
in forward Compton scattering where the usual structure functions
$F_{1}$ and $F_{2}$ are calculated for both spinless and
spin-$\frac{1}{2}$ hadrons. An infrared cutoff in the fifth
dimension at $z=z_{0}\sim 1/\Lambda _{QCD}$, which breaks the
conformal symmetry in CFT, is introduced to give rise to
confinement. In addition, this cutoff scale also provides a mass
scale for the hadrons with $M\propto 1/z_{0}$. This particular
model with the hard cutoff in the infrared is then called the hard
wall model. There have been a lot of interesting
studies\cite{Brower:2002er, BoschiFilho:2002zs} and further
developments\cite{Brower:2007xg, Brower:2006ea,
BallonBayona:2008zi, BallonBayona:2007rs, BallonBayona:2007qr,
Pire:2008zf, Cornalba:2008sp} along this direction. These studies
of DIS and structure functions have provided an interesting
picture of hadrons at the strong coupling limit. One finds that there
is no large-$x$ parton which carries a finite amount of longitudinal
momentum of the target hadron and most of the longitudinal
momentum is distributed in the partons sitting in the extremely
low-$x$ region. Furthermore, DIS, in terms of dipole
(quark-antiquark pair with an open string attached) scattering on
a hadron/nucleus target, is discussed in
Refs.~\cite{Albacete:2008ze, Levin:2009vj, Kovchegov:2009yj}. In
particular, the AdS/CFT computation of deep inelastic scattering
off the finite temperature plasma has been recently studied in
Refs.~\cite{Hatta:2007cs,
Mueller:2008bt, Avsar:2009xf, Dominguez:2009cm, Bayona:2009qe, Iancu:2009py}%
, which also indicate that most of the constituents in the strongly coupled
plasma are located in the very low $x$ region. Based on hard scatterings in
AdS/CFT, the saturation picture is then developed in Refs.~\cite%
{Hatta:2007he, Dominguez:2008vd}.

Recently, we extended the above DIS calculation in AdS/CFT to the case of
polarized DIS in Ref.~\cite{Gao:2009ze} and obtained the spin-dependent
structure functions $g_{1}$ and $g_{2}$ for a spin-$\frac{1}{2}$ hadron. In
Ref.~\cite{Hatta:2009ra}, the spin decomposition of spin-$\frac{1}{2}$
hadrons at large coupling limit was analyzed. These studies show that the
spin of hadrons at large t'Hooft coupling seems solely due to orbital
angular momentum. Suggested by Ji\cite{Ji:1996ek} in QCD, the deeply virtual
Compton scattering (DVCS) can be used as the gateway to the total quark and
gluon contributions to the spin of the proton. The DVCS amplitude contains
the important information about the total spin of constituents in the hadron
which is encoded in the generalized parton distributions. In addition, one
can access the orbital angular momentum of quarks and gluons by subtracting
the spin contributions. Therefore, in order to further investigate the spin
and the orbital angular momentum, we need to study the nonforward Compton
scattering and extract the so-called generalized parton distributions\cite%
{Mueller:1998fv, Ji:1996ek, Ji:1996nm, Radyushkin:1996nd, Radyushkin:1997ki,
Ji:2004gf, Belitsky:2005qn, Goeke:2001tz, Diehl:2003ny, Brodsky:2000xy}. As
a first step, we study the nonforward Compton scattering, in particular, the
DVCS, in the AdS/CFT correspondence. We will leave the extraction of
generalized parton distributions for future study.

The objective of this paper is to study the nonforward Compton scattering.
We use dilatons (scalar) or dilatinos (fermion) as the targets which are
scattered by a virtual gauge boson field (a physical virtual photon, so to
speak). We are particularly interested in the DVCS since it is widely
studied in QCD and of great importance in spin physics. The final states
contain a photon, which is real in DVCS, and the dilaton or dilatino target.
This calculation is equivalent to a four-point function calculation that
involves two currents $J$ and two targets (dilaton or dilatino) in the
AdS/CFT. Four-point functions of the dilaton and axion are studied in Refs.~\cite%
{Liu:1998ty, Freedman:1998bj, D'Hoker:1999pj}. Recently, four-point
functions of $\mathcal{R}$-currents are discussed in AdS supergravity\cite%
{Bartels:2009sc}.

To be more specific, we compute the supergravity graphs corresponding to
Compton scattering in CFT. For the case of the dilaton, we have calculated the
\textit{s}-channel and \textit{u}-channel graphs where all the Kaluza-Klein excitations of
intermediate states are summed over. We have also included a four-point
contact term and shown that the sum of these three graphs are gauge
invariant, namely, satisfying the current conservation constraint. For the
case of the dilatino, we compute the contributions from the \textit{s}-channel and
\textit{u}-channel graphs explicitly. Assuming that protons can be identified as
supergravity modes of the dilatino, we compare the calculated DVCS cross section
to the low-energy H1 and ZEUS data, and find that our calculation is
consistent with the data. We also study the real Compton scattering and find
that the scattering amplitudes are parametrically the same as the one found
in QED.

We notice that there is an additional \textit{t}-channel graviton exchange contribution,
which is supposed to be dominant at high energy. We provide a heuristic
derivation for the computation of \textit{t}-channel graviton exchange at the end of
this paper. Since the graviton exchange only couples to the energy momentum
tensor of the target, we find the results for the dilaton and dilatino are quite
similar.

The rest  of the paper is organized in the following manner: In
Sec.\ref{scalar}, we consider the usual \textit{s}-channel and \textit{u}-channel
exchanges as well as the four-point interaction contributions for a dilaton
target. In Sec.\ref{fermion}, we evaluate the supergravity graphs of the
\textit{s}-channel and \textit{u}-channel exchanges for dilatino targets. The \textit{t}-channel
exchange of the graviton for both dilaton and dilatino targets is discussed in
Sec.\ref{gravitonex}. We summarize and present some more discussion in
Sec.\ref{conclusion}.

\section{Compton scattering of the dilaton}

\label{scalar} In this section, we will formulate the Compton scattering on
a scalar target in the CFT.

\subsection{Scaler field in AdS space and its bulk to bulk propagator}

First, let us consider the free scalar field and derive its bulk to bulk
propagator in $\textrm{AdS}_{5}$ space. Scalars in CFT correspond to supergravity
modes of dilatons. The metric in $\textrm{AdS}\times S^{5}$ space can be written as
\begin{equation}
ds^{2}=\frac{R^{2}}{z^{2}}(\eta _{\mu \nu }dy^{\mu }dy^{\nu }+dz^{2})+R^{2}%
\text{d}\Omega _{5}^{2},
\end{equation}%
where $\eta _{\mu \nu }=\left( -,+,+,+\right) $ is the mostly plus flat
space metric. We assume that the initial/final dilaton wave function can be
written as
\begin{equation}
\Phi (y,z)=e^{ip\cdot y}\phi (z)
\end{equation}%
where, for simplicity, we have neglected the dependence on the coordinates $%
S^{5}$. The equation of motion of the dilaton in AdS space can be written as%
\begin{equation}
\frac{1}{\sqrt{-g}}\partial _{m}\left( \sqrt{-g}g^{mn}\partial _{n}\Phi
(y,z)\right) -\mu ^{2}\Phi (y,z)=0,
\end{equation}%
with the fifth dimension mass $\mu ^{2}=\frac{\Delta (\Delta -4)}{R^{2}}$.
It reduces to
\begin{equation}
z^{2}\partial _{z}^{2}\Phi (y,z)-3z\partial _{z}\Phi (y,z)+z^{2}\Box \Phi
(y,z)-\Delta (\Delta -4)\Phi (y,z)=0,
\end{equation}%
where $\Box =-\partial _{t}^{2}+\nabla ^{2}$. The solution to the above equation
is given by
\[
\phi (z)=C_{1}z^{2}J_{\Delta -2}(Mz)+C_{2}z^{2}Y_{\Delta -2}(Mz),
\]%
where $M^{2}=-p^{2}$. With the consideration of boundary conditions for
normalizable modes, we set $C_{2}=0$ and choose only the $J_{\Delta -2}(Mz)$
part.

In order to calculate the Compton scattering amplitude, we also need the
bulk to bulk propagator of dilatons in $A\textrm{d}S_{5}$ space, which satisfies
\begin{equation}
\left[ z^{2}\partial _{z}^{2}-3z\partial _{z}+z^{2}\Box -\Delta (\Delta -4)%
\right] G(x,z;y,z^{\prime })=z^{5}\delta (z-z^{\prime })\delta ^{(4)}(x-y).
\label{green}
\end{equation}%
In $\textrm{AdS}_{5}$ space, from coordinate $(x,z)$ to $(y,z^{\prime })$, the
propagator can be written as \cite{Liu:1998ty, Avsar:2009xf},
\begin{equation}
G(x,z;y,z^{\prime })=-\int \frac{d^{4}k}{(2\pi )^{4}}e^{-ik\cdot
(x-y)}\int_{0}^{\infty }d\omega \frac{\omega }{\omega ^{2}+k^{2}-i\epsilon }%
z^{2}J_{\Delta -2}(\omega z)z^{\prime 2}J_{\Delta -2}(\omega z^{\prime }),
\end{equation}%
where $x$ and $y$ are four-dimensional coordinates. One can easily check
the above expression satisfies Eq.~(\ref{green}) by noticing that
\begin{equation}
\left[ z^{2}\partial _{z}^{2}-3z\partial _{z}-\Delta (\Delta -4)\right]
z^{2}J_{\Delta -2}(\omega z)=-\omega ^{2}z^{2}J_{\Delta -2}(\omega z)
\end{equation}%
and
\begin{equation}
\int_{0}^{\infty }d\omega \omega z^{2}J_{\Delta -2}(\omega z)z^{\prime
2}J_{\Delta -2}(\omega z^{\prime })=z^{3}\delta (z-z^{\prime }).
\label{norm}
\end{equation}

Note that $k^{2}<0$ corresponds to the timelike case, and $k^{2}>0$
corresponds to spacelike case. When the particle is at rest, we shall have $%
k^{2}=-M^{2}$. For the case of the \textit{s}-channel Compton scattering, we find
that $k=p+q=p^{\prime }+q^{\prime }$. The dilaton propagator in the fifth
dimension is then defined as
\begin{equation}
G(z;z^{\prime })=-\int_{0}^{\infty }d\omega \frac{\omega }{\omega
^{2}-M_{X}^{2}-i\epsilon }z^{2}J_{\Delta -2}(\omega z)z^{\prime 2}J_{\Delta
-2}(\omega z^{\prime }),
\end{equation}%
with $M_{X}^{2}=-(p+q)^{2}$ for the \textit{s}-channel scattering and $%
M_{X}^{2}=-(p-q^{\prime })^{2}$ for the \textit{u}-channel scattering.

\subsection{Kaluza-Klein field in AdS Space}

Following the prescription in Ref. \cite{Polchinski:2002jw}, the
incident/outgoing current is chosen to be the $\mathcal{R}$ current which
couples to the hadron as an isometry of $S^{5}$ with the killing vector $v_a$.
This is equivalent to saying that the current couples to the target with a conserved charge $\mathcal{Q}$. According to the AdS/CFT
correspondence, the current excites a nonnormalizable mode of a Kaluza-Klein
gauge field on the Minkowski boundary of the $AdS_{5}$ space
\begin{equation}
\delta G_{ma}=A_{m}(y,r)v_{a}(\Omega ),
\end{equation}%
This gauge field fluctuation $A_{m}(y,r)$ (a physical virtual photon, so to
speak) can be viewed as a vector boson field which couples to the $\mathcal{R%
}$ current $J^{\mu }$ on the Minkowski boundary, and then propagates into
the bulk as gravitational waves, and eventually interacts with the
supergravity modes of the dilaton or dilatino. The gauge field satisfies
Maxwell's equation in the bulk, $D_{m}F^{mn}=0$ which can be explicitly
written as
\begin{equation}
\frac{1}{\sqrt{-g}}\partial _{m}\left[ \sqrt{-g}g^{nk}g^{ml}\left( \partial
_{k}A_{l}-\partial _{l}A_{k}\right) \right] =0,
\end{equation}%
where $m$, $n$, ... are indices on $AdS_{5}$. With the boundary condition
\begin{equation}
A_{\mu }(y,\infty )=A_{\mu }(y)|_{\mathrm{4d}}=n_{\mu }e^{iq\cdot y}.
\end{equation}%
and the Lorentz-like gauge
\begin{equation}
\partial _{\mu }A^{\mu }+z\partial _{z}\left( \frac{A_{z}}{z}\right) =0,
\end{equation}%
the Maxwell equation can be written as
\begin{eqnarray}
-q^{2}A_{\mu }+z\partial _{z}\left( \frac{1}{z}\partial _{z}A_{\mu }\right)
&=&0, \\
-q^{2}A_{z}+\partial _{z}\left( z\partial _{z}\left( \frac{1}{z}A_{z}\right)
\right) &=&0.
\end{eqnarray}%
The solutions to the above equations are given by,
\begin{eqnarray}
A_{\mu } &=&n_{\mu }e^{iq\cdot y}qz{K}_{1}(qz),  \nonumber \\
A_{z} &=&in\cdot qe^{iq\cdot y}z{K}_{0}(qz),
\end{eqnarray}%
where ${K}_{1}$ and ${K}_{0}$ are both modified Bessel functions and $n_{\mu
}$ is the polarization vector. Note that $q^2=Q^2>0$ for the spacelike current in the mostly plus metric signature.

\subsection{Compton scattering amplitudes}

\begin{figure}[tbp]
\begin{center}
\includegraphics[width=6cm]{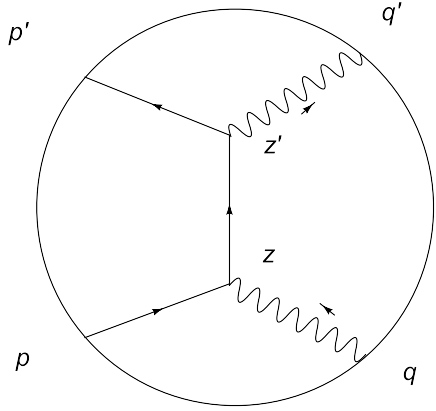}
\end{center}
\caption[*]{\textit{s}-channel Compton scattering}
\label{DVCS}
\end{figure}
\

\begin{figure}[tbp]
\begin{center}
\includegraphics[width=6cm]{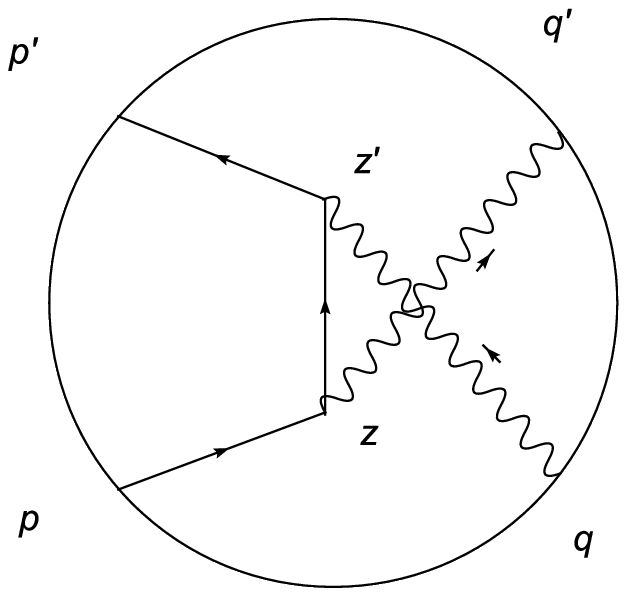}
\end{center}
\caption[*]{\textit{u}-channel Compton scattering}
\label{DVCS2}
\end{figure}

\begin{figure}[tbp]
\begin{center}
\includegraphics[width=6cm]{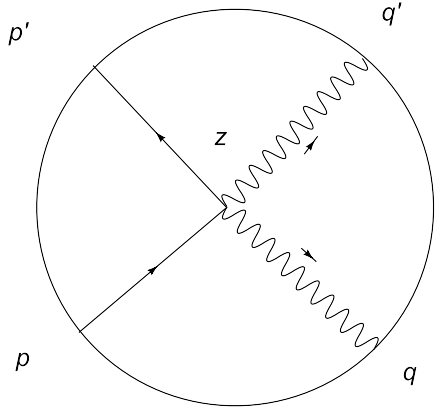}
\end{center}
\caption[*]{Four-point interaction of Compton scattering}
\label{DVCS3}
\end{figure}

Now we are ready to consider the Compton scattering on a dilaton target. The
relevant supergravity graphs are shown in Figs.~\ref{DVCS}--\ref{DVCS3}, corresponding to $s$-channel exchange, $u$-channel
exchange and four-point interaction, respectively. The $t$-channel graviton
exchange illustrated in Fig.~\ref{DVCS4} will be considered in Sec.\ref%
{gravitonex}.

The standard Compton amplitude is defined as
\begin{eqnarray}
T_{\mu\nu}=i\int d^4y e^{-iq\cdot y}\langle P^{\prime}|\text{T} J_\mu(y)
J_\nu(0)|P\rangle.
\end{eqnarray}

We first start with the calculation of the \textit{s}-channel amplitude as
illustrated in Fig.~\ref{DVCS}. The bulk to bulk propagator from $z$ to $%
z^{\prime }$ represents high excited states of the dilaton target after
absorbing the incoming virtual photon. This is different from the DIS
calculation where the propagator is always put on shell. The relevant
supergravity interaction of the dilaton and photon is
\begin{eqnarray}
S_{int} &=&\int d^{10}x\sqrt{-g}A^{m}v^{a}\partial _{m}\Phi \partial _{a}\Phi
\nonumber \\
&=&i{\mathcal{Q}}\int d^{10}x\sqrt{-g}A^{m}(\Phi \partial _{m}\Phi ^{\ast
}-\Phi ^{\ast }\partial _{m}\Phi ).
\end{eqnarray}%
The dilaton is taken to be in a charge eigenstate with the charge $\mathcal{Q%
}$ under the $U(1)$ symmetry. This yields $v_{a}\partial ^{a} \Phi =i%
\mathcal{Q}\Phi$.

Using the dilaton wave functions along with the bulk to bulk propagator and
above interaction vertex, it is straightforward to write down the \textit{s}-channel
amplitude
\begin{equation}
n^{\mu }T_{\mu \nu }^{s}n^{\prime \ast \nu }=\frac{R^{10}}{R^{8}}\int
d^{4}xd^{4}ydzdz^{\prime }\Phi _{i}(x,z)n^{\mu }J_{\mu }(z)G(x,z;y,z^{\prime
})n^{\prime \ast \nu }J_{\nu }^{\ast }(z^{\prime })\Phi _{f}^{\ast
}(y,z^{\prime }),
\end{equation}%
where each integration over the $S^{5}$ space gives a factor of $R^{5}$ and
the Green's function $G(x,z;y,z^{\prime })$ picks up a factor of $\frac{1}{%
R^{8}}$ when we take the $S^{5}$ space into account. (One can actually set $%
R=1$ in the calculation since the final result should not depend on the
curvature $R$.) The initial and final wave functions of the dilaton are
\begin{equation}
\Phi _{i}(x,z)=\frac{c_{i}}{z_{0}R^{4}}z^{2}J_{\Delta -2}(Mz)e^{ip\cdot
x}\quad \text{and}\quad \Phi _{f}^{\ast }(y,z)=\frac{c_{f}}{z_{0}R^{4}}%
z^{2}J_{\Delta -2}(Mz)e^{-ip^{\prime }\cdot y},
\end{equation}%
respectively. The boundary condition is set as $J_{\Delta -2}(Mz_{0})=0$ in
the fifth dimension to mimic confinement. In addition, this cutoff scale
also yields a mass scale for the hadrons with $M\propto 1/z_{0}$\cite%
{deTeramond:2005su, Brodsky:2006uqa}. It is straightforward to integrate out
$x$ and $y$, and then find
\begin{equation}
n^{\mu }T_{\mu \nu }^{s}n^{\prime \ast \nu }=(2\pi )^{4}\delta
^{(4)}(p+q-p^{\prime }-q^{\prime })R^{2}\int dzdz^{\prime }\Phi
_{i}(z)n^{\mu }J_{\mu }(z)G(z;z^{\prime })n^{\prime \ast \nu }J_{\nu }^{\ast
}(z^{\prime })\Phi _{f}^{\ast }(z^{\prime }),
\end{equation}%
The \textit{s}-channel amplitude $T_{\mu \nu }^{s}$ can be cast into
\[
T_{\mu \nu }^{s}\propto (2p_{\mu }+q_{\mu })(2p_{\nu }^{\prime }+q_{\nu
}^{\prime }){\mathcal{T}}_{1}^{s}-(2p_{\mu }+q_{\mu })q_{\nu }^{\prime }{%
\mathcal{T}}_{2}^{s}-q_{\mu }(2p_{\nu }^{\prime }+q_{\nu }^{\prime }){%
\mathcal{T}}_{3}^{s}+q_{\mu }q_{\nu }^{\prime }{\mathcal{T}}_{4}^{s},
\]%
where
\[
{\mathcal{T}}_{1}^{s}=\int dzdz^{\prime }qJ_{\Delta
-2}(Mz)K_{1}(qz)G(z,z^{\prime })q^{\prime }J_{\Delta -2}(Mz^{\prime
})K_{1}(q^{\prime }z^{\prime })
\]%
\begin{eqnarray}
{\mathcal{T}}_{2}^{s} &=&\int dzdz^{\prime }qJ_{\Delta
-2}(Mz)K_{1}(qz)\partial _{z^{\prime }}G(z,z^{\prime })J_{\Delta
-2}(Mz^{\prime })K_{0}(q^{\prime }z^{\prime })  \nonumber \\
&&-\int dzdz^{\prime }qJ_{\Delta -2}(Mz)K_{1}(qz)G(z,z^{\prime })\frac{1}{%
z^{\prime 2}}\partial _{z^{\prime }}\left[ z^{\prime 2}J_{\Delta
-2}(Mz^{\prime })\right] K_{0}(q^{\prime }z^{\prime }),
\end{eqnarray}%
\begin{eqnarray}
{\mathcal{T}}_{3}^{s} &=&\int dzdz^{\prime }qJ_{\Delta
-2}(Mz)K_{0}(qz)\partial _{z}G(z,z^{\prime })J_{\Delta -2}(Mz^{\prime
})K_{1}(q^{\prime }z^{\prime })  \nonumber \\
&&-\int dzdz^{\prime }\frac{1}{z^{2}}\partial _{z}\left[ z^{2}J_{\Delta
-2}(Mz)\right] K_{0}(qz)G(z,z^{\prime })q^{\prime }J_{\Delta -2}(Mz^{\prime
})K_{1}(q^{\prime }z^{\prime }),
\end{eqnarray}%
and
\begin{eqnarray}
{\mathcal{T}}_{4}^{s} &=&\int dzdz^{\prime }J_{\Delta
-2}(Mz)K_{0}(qz)\partial _{z^{\prime }}\partial _{z}G(z,z^{\prime
})J_{\Delta -2}(Mz^{\prime })K_{0}(q^{\prime }z^{\prime })  \nonumber \\
&&-\int dzdz^{\prime }J_{\Delta -2}(Mz)K_{0}(qz)\partial _{z}G(z,z^{\prime })%
\frac{1}{z^{\prime 2}}\partial _{z^{\prime }}\left[ z^{\prime 2}J_{\Delta
-2}(Mz^{\prime })\right] K_{0}(q^{\prime }z^{\prime })  \nonumber \\
&&-\int dzdz^{\prime }K_{0}(qz)\frac{1}{z^{2}}\partial _{z}\left[
z^{2}J_{\Delta -2}(Mz)\right] \partial _{z^{\prime }}G(z,z^{\prime
})J_{\Delta -2}(Mz^{\prime })K_{0}(q^{\prime }z^{\prime })  \nonumber \\
&&+\int dzdz^{\prime }K_{0}(qz)\frac{1}{z^{2}}\partial _{z}\left[
z^{2}J_{\Delta -2}(Mz)\right] G(z,z^{\prime })\frac{1}{z^{\prime 2}}\partial
_{z^{\prime }}\left[ z^{\prime 2}J_{\Delta -2}(Mz^{\prime })\right]
K_{0}(q^{\prime }z^{\prime }).
\end{eqnarray}

Using the Eq.~(\ref{norm}) and the recurrence relations
\begin{eqnarray}
&&K_{0}(z)=-\frac{d}{dz}K_{1}(z)-\frac{K_{1}(z)}{z},  \label{kr} \\
&&J_{\nu }(\alpha z)\frac{d}{dz}\left[ z\frac{d}{dz}J_{\nu }(\beta z)\right]
-J_{\nu }(\beta z)\frac{d}{dz}\left[ z\frac{d}{dz}J_{\nu }(\alpha z)\right]
=(\alpha ^{2}-\beta ^{2})zJ_{\nu }(\alpha z)J_{\nu }(\beta z),
\end{eqnarray}%
one obtains
\begin{eqnarray}
&&\int dz^{\prime }\left\{ \partial _{z^{\prime }}G(z,z^{\prime })J_{\Delta
-2}(Mz^{\prime })-\frac{G(z,z^{\prime })}{z^{\prime 2}}\partial _{z^{\prime
}}\left[ z^{\prime 2}J_{\Delta -2}(Mz^{\prime })\right] \right\}
K_{0}(q^{\prime }z^{\prime })  \nonumber \\
&=&\frac{M^{2}+s}{q^{\prime 2}}\int dz^{\prime }q^{\prime }G(z,z^{\prime
})J_{\Delta -2}(Mz^{\prime })K_{1}(q^{\prime }z^{\prime
})+z^{3}K_{1}(q^{\prime }z)J_{\Delta -2}(Mz)/q^{\prime },
\end{eqnarray}%
where $s=(p+q)^{2}$. Eventually, one can simplify and cast these $\mathcal{T%
}$ amplitudes into
\begin{eqnarray}
{\mathcal{T}}_{2}^{s} &=&\left( 1-\frac{1}{x^{\prime }}\right) \mathcal{T}%
_{1}^{s}+\frac{q}{q^{\prime }}\int dzz^{3}J_{\Delta
-2}^{2}(Mz)K_{1}(qz)K_{1}(q^{\prime }z), \\
{\mathcal{T}}_{3}^{s} &=&\left( 1-\frac{1}{x}\right) \mathcal{T}_{1}^{s}+%
\frac{q^{\prime }}{q}\int dzz^{3}J_{\Delta
-2}^{2}(Mz)K_{1}(qz)K_{1}(q^{\prime }z),
\end{eqnarray}%
and
\begin{eqnarray}
{\mathcal{T}}_{4}^{s} &=&\left( \frac{1}{x^{\prime }}-1\right) \left( \frac{1%
}{x}-1\right) \mathcal{T}_{1}^{s}-\int dzz^{3}J_{\Delta
-2}^{2}(Mz)K_{0}(qz)K_{0}(q^{\prime }z)  \nonumber \\
&&+\frac{M^{2}+s}{q^{\prime }q}\int dzz^{3}J_{\Delta
-2}^{2}(Mz)K_{1}(qz)K_{1}(q^{\prime }z),
\end{eqnarray}%
where
\[
x=-\frac{q^{2}}{2p\cdot q},\ \ \ x^{\prime }=-\frac{q^{\prime 2}}{2p^{\prime
}\cdot q^{\prime }}.
\]%
In the end, we obtain the \textit{s}-channel amplitude
\begin{eqnarray}
T_{\mu \nu }^{s} &\propto &\left( 2p_{\mu }+\frac{q_{\mu }}{x}\right) \left(
2p_{\nu }^{\prime }+\frac{q_{\nu }^{\prime }}{x^{\prime }}\right) \mathcal{T}%
_{1}^{s}-q_{\mu }q_{\nu }^{\prime }\mathcal{C}_{0}  \nonumber \\
&&-\left[ \left( 2p_{\mu }+q_{\mu }\right) q_{\nu }^{\prime }\frac{q}{%
q^{\prime }}+q_{\mu }\left( 2p_{\nu }^{\prime }+q_{\nu }^{\prime }\right)
\frac{q^{\prime }}{q}-q_{\mu }q_{\nu }^{\prime }\frac{M^{2}+s}{qq^{\prime }}%
\right] \mathcal{C}_{1},
\end{eqnarray}%
where we have defined that
\begin{eqnarray}
\mathcal{C}_{0} &=&\int dzz^{3}J_{\Delta -2}^{2}(Mz)K_{0}(qz)K_{0}(q^{\prime
}z) \\
\mathcal{C}_{1} &=&\int dzz^{3}J_{\Delta -2}^{2}(Mz)K_{1}(qz)K_{1}(q^{\prime
}z).
\end{eqnarray}%
Similarly, the amplitude from the \textit{u}-channel diagram, which is illustrated in Fig.~\ref{DVCS2}, is found to be
\begin{eqnarray}
T_{\mu \nu }^{u} &\propto &\left( 2p_{\mu }^{\prime }-\frac{q_{\mu }}{\tilde{%
x}}\right) \left( 2p_{\nu }-\frac{q_{\nu }^{\prime }}{\tilde{x}^{\prime }}%
\right) \mathcal{T}_{1}^{u}-q_{\mu }q_{\nu }^{\prime }\mathcal{C}_{0}
\nonumber \\
&&+\left[ \left( 2p_{\mu }^{\prime }-q_{\mu }\right) q_{\nu }^{\prime }\frac{%
q}{q^{\prime }}+q_{\mu }\left( 2p_{\nu }-q_{\nu }^{\prime }\right) \frac{%
q^{\prime }}{q}+q_{\mu }q_{\nu }^{\prime }\frac{M^{2}+u}{qq^{\prime }}\right]
\mathcal{C}_{1},
\end{eqnarray}%
where
\begin{equation}
\tilde{x}=\frac{q^{2}}{2p^{\prime }\cdot q},\quad \tilde{x}^{\prime }=\frac{%
q^{\prime 2}}{2p\cdot q^{\prime }},
\end{equation}%
and $u=(p^{\prime }-q)^{2}$. Note that actually the \textit{u}-channel amplitude and
\textit{s}-channel amplitude are related by interchanging $q_{\mu },n^{\mu }$ and $%
q_{\nu }^{\prime },n^{\prime \ast \nu }$ according to the crossing symmetry.

Last but not least, we need to calculate the four-point interaction
contribution, namely, the contact term as depicted in Fig.~\ref{DVCS3}. The four-point interaction amplitude
is
\begin{equation}
n^{\mu }T_{\mu \nu }^{c}n^{\prime \ast \nu }=2\mathcal{Q}^{2}R^{5}(2\pi
)^{4}\delta ^{(4)}(p+q-p^{\prime }-q^{\prime })\int dz\sqrt{-g}\Phi
_{i}(z)g^{mn}A_{m}A_{n}^{\ast }\Phi _{f}^{\ast }(z).
\end{equation}%
Then it is easy to obtain
\begin{eqnarray}
T_{\mu \nu }^{c} &\propto &2\eta _{\mu \nu }q^{\prime }q\int
dzz^{3}J_{\Delta -2}^{2}(Mz)K_{1}(qz)K_{1}(q^{\prime }z)  \nonumber \\
&&+2q_{\mu }q_{\nu }^{\prime }\int dzz^{3}J_{\Delta
-2}^{2}(Mz)K_{0}(qz)K_{0}(q^{\prime }z).
\end{eqnarray}

Summing over all the contributions from those supergravity graphs, we arrive
at the gauge invariant results
\begin{eqnarray}
T_{\mu \nu }^{\mathrm{tot}} &=&T_{\mu \nu }^{s}+T_{\mu \nu }^{u}+T_{\mu \nu }^{c}
\nonumber \\
&=&\mathcal{A}\left[ \left( 2p_{\mu }+\frac{q_{\mu }}{x}\right) \left(
2p_{\nu }^{\prime }+\frac{q_{\nu }^{\prime }}{x^{\prime }}\right) \mathcal{T}%
_{1}^{s}+\left( 2p_{\mu }^{\prime }-\frac{q_{\mu }}{\tilde{x}}\right) \left(
2p_{\nu }-\frac{q_{\nu }^{\prime }}{\tilde{x}^{\prime }}\right) \mathcal{T}%
_{1}^{u}\right]  \nonumber \\
&&+2\mathcal{A}\left( \eta _{\mu \nu }q^{\prime }q-q_{\mu }^{\prime }q_{\nu
}^{\prime }\frac{q}{q^{\prime }}-q_{\mu }q_{\nu }\frac{q^{\prime }}{q}%
+q_{\mu }q_{\nu }^{\prime }\frac{q\cdot q^{\prime }}{q^{\prime }q}\right)
\mathcal{C}_{1}.
\end{eqnarray}%
The overall constant is found to be $\mathcal{A}=\frac{\mathcal{Q}%
^{2}c_{i}c_{f}}{z_{0}^{2}}(2\pi )^{4}\delta ^{(4)}(p+q-p^{\prime }-q^{\prime
})$. One can easily check the conservation of the current
\begin{equation}
q^{\mu }T_{\mu \nu }^{\mathrm{tot}}=0\quad \text{and}\quad T_{\mu \nu
}^{tot}q^{\prime \nu }=0.
\end{equation}

It is well-known that the DIS amplitude and structure functions can be
extracted from the imaginary part of forward Compton scattering amplitude
according to the optical theorem. Therefore, one can easily obtain the
structure functions $F_{1}$ and $F_{2}$ by taking the forward limit and the
imaginary part of $T_{\mu \nu }^{\mathrm{tot}}$. The results agree with those found
in Ref. \cite{Polchinski:2002jw} in the large-$x$ region. The same
conclusion can be also drawn for a dilatino target.

\subsection{Real Compton scattering}

In the case of real Compton scattering, it follows that $q^2=q^{\prime 2 }=0$
since the incoming and outgoing photons are taken to be real photons. Since
a real photon only has 2 degrees of freedom, one can choose a polarization
vector $\varepsilon ^{\mu }\left( q,\lambda \right) $ which is orthogonal to
$p_{\mu }$ and $q_{\mu }$ in a special gauge. Similarly we can also choose
to have $\varepsilon ^{\prime \nu }\left( q^{\prime },\lambda ^{\prime
}\right) q_{\nu }^{\prime }=0$ and $\varepsilon ^{\prime \nu }\left(
q^{\prime },\lambda ^{\prime }\right) p_{\nu }=0$ for the outgoing photon in
the target rest frame. Therefore, the amplitude of real Compton scattering
can be reduced to
\begin{equation}
\mathcal{M}=2\mathcal{Q}^{2}c_{i}c_{f}\varepsilon \left( q,\lambda \right)
\cdot \varepsilon ^{\prime }\left( q^{\prime },\lambda ^{\prime }\right) B,
\end{equation}%
with $B=\frac{1}{z_{0}^{2}}\int_{0}^{z_{0}}dz\,zJ_{\Delta -2}^{2}(Mz)=\frac{1%
}{2}\left[ J_{\Delta -2}^{2}(Mz_{0})-J_{\Delta -1}(Mz_{0})J_{\Delta
-3}(Mz_{0})\right] >0$ when $z_{0}>0$.
This eventually leads to the
unpolarized Compton cross section after averaging over $\lambda $ and
summing over $\lambda ^{\prime }$ which gives $\frac{1}{2}%
\sum\limits_{\lambda ,\lambda ^{\prime }}\left\vert \varepsilon \left(
q,\lambda \right) \cdot \varepsilon ^{\prime }\left( q^{\prime },\lambda
^{\prime }\right) \right\vert ^{2}=\frac{1}{2}\left( 1+\cos ^{2}\theta
\right) $. Thus eventually one gets
\begin{equation}
\frac{d\sigma ^{\text{Un}}}{d\Omega _{q^{\prime }}}\varpropto \mathcal{Q}^{4}%
\frac{q_{0}^{\prime 2}}{q_{0}^{2}}\left( 1+\cos ^{2}\theta \right)
,\label{real}
\end{equation}%
where $\theta $ is the angle between the incoming photon and
outgoing photon. This result\footnote{The real Compton amplitude
is sensitive to the IR region of the target wave function.
 For a different geometry other than AdS, we expect that  Eq.~(\ref{real}) will still hold while the constant
$B$ may change since it depends on the details of the IR cutoff.}
is identical to the cross section found in scalar electrodynamics
except for the overall constant.\footnote{ Here we do not take the
\textit{t}-channel graviton exchange contribution into account since it
belongs to another class of contributions. The discussion on this
issue is given in the end of Sec.~\ref{fermion} and the
calculation
of \textit{t}-channel graviton exchange amplitude is provided in Sec.~\ref{gravitonex}%
.}

\subsection{Deeply virtual Compton scattering}

In the case of deeply virtual Compton scattering, only the final state
photon is real. Therefore, we set $q^{\prime 2}=0$ but assume $q^{2}$ is
large as compared to \thinspace $M^{2}$. Hereafter in this subsection, we
work in a limit that $M\rightarrow 0$ and $z_{0}\rightarrow \infty $ but
with $Mz_{0}$ fixed. In this case, the outgoing real photon has 2 degrees
of freedom, while the incoming virtual photon has 3. It is usually
convenient to choose to stay in the target rest frame (the lab frame). In
this frame, we can choose to have $\varepsilon ^{\prime \nu }\left(
q^{\prime },\lambda ^{\prime }\right) q_{\nu }^{\prime }=0$ and $\varepsilon
^{\prime \nu }\left( q^{\prime },\lambda ^{\prime }\right) p_{\nu }=0$ as
well as $\varepsilon ^{\mu }\left( q,\lambda \right) q_{\mu }=0$. This
yields
\begin{equation}
\mathcal{M}=2\mathcal{Q}^{2}\frac{c_{i}c_{f}}{z_{0}^{2}}\varepsilon \left(
q,\lambda \right) \cdot \varepsilon ^{\prime }\left( q^{\prime },\lambda
^{\prime }\right) \mathcal{C}_{1\text{DVCS}}+4\mathcal{Q}^{2}\frac{c_{i}c_{f}%
}{z_{0}^{2}}\varepsilon \left( q,\lambda \right) \cdot p\varepsilon ^{\prime
}\left( q^{\prime },\lambda ^{\prime }\right) \cdot p^{\prime }\mathcal{T}_{1%
\text{DVCS}}^{s},
\end{equation}%
where%
\begin{eqnarray}
\mathcal{C}_{1\text{DVCS}} &=&\int_{0}^{z_{0}}dzz^{2}J_{\Delta
-2}^{2}(Mz)qK_{1}(qz) \\
&\simeq &\frac{2\left( \Delta -1\right) }{M^{2}}\left( \frac{M^{2}}{q^{2}}%
\right) ^{\Delta -1}\left. _{2}F_{1}\right. (\Delta -\frac{3}{2},\Delta
,2\Delta -3,-\frac{4M^{2}}{q^{2}})
\end{eqnarray}%
and%
\begin{eqnarray}
\mathcal{T}_{1\text{DVCS}}^{s} &=&-\int_{0}^{\infty }d\omega \frac{\omega }{%
\omega ^{2}+(p+q)^{2}-i\epsilon }\int_{0}^{z_{0}}dzqJ_{\Delta
-2}(Mz)K_{1}(qz)z^{2}J_{\Delta -2}(\omega z)  \nonumber \\
&&\times \int_{0}^{z_{0}}dz^{\prime }z^{\prime }J_{\Delta -2}(\omega
z^{\prime })J_{\Delta -2}(Mz^{\prime })  \nonumber \\
&\simeq &\frac{-1}{M^{2}+(p+q)^{2}-i\epsilon }\frac{2\left( \Delta -1\right)
}{M^{2}}\left( \frac{M^{2}}{q^{2}}\right) ^{\Delta -1}\left.
_{2}F_{1}\right. (\Delta -\frac{3}{2},\Delta ,2\Delta -3,-\frac{4M^{2}}{q^{2}%
}),
\end{eqnarray}%
where we approximately write $\int_{0}^{z_{0}}dz^{\prime }z^{\prime
}J_{\Delta -2}(\omega z^{\prime })J_{\Delta -2}(Mz^{\prime })\simeq \frac{1}{%
M}\delta (\omega -M)$, which is an exact result if $z_{0}\rightarrow \infty $%
. One can approximately write $\mathcal{T}_{1\text{DVCS}}^{s}\simeq \frac{-1%
}{M^{2}+(p+q)^{2}-i\epsilon }\mathcal{C}_{1\text{DVCS}}$.

To obtain the cross section, let us work out the kinematics first. We
analyze the deeply virtual Compton scattering in the lab frame, in which the
initial scalar is at rest. Thus,
\begin{eqnarray}
\text{Before the scattering }p_{\mu } &=&\left( M,0,0,0\right) ,\text{ \ }%
q_{\mu }=\left( q_{0},0,0,q_{3}\right) ;  \label{frame1} \\
\text{After the scattering }p_{\mu }^{\prime } &=&\left( p_{0}^{\prime
},p_{1}^{\prime },0,p_{3}^{\prime }\right) ,\text{ }q_{\mu }^{\prime
}=\left( \omega ^{\prime },\omega ^{\prime }\sin \theta ,0,\omega ^{\prime
}\cos \theta \right) .  \label{frame2}
\end{eqnarray}%
According to the conservation of energy momentum, it is straightforward to
see that
\begin{eqnarray}
p_{0}^{\prime } &=&M+q_{0}-\omega ^{\prime }, \\
p_{1}^{\prime } &=&-\omega ^{\prime }\sin \theta , \\
p_{3}^{\prime } &=&q_{3}-\omega ^{\prime }\cos \theta .
\end{eqnarray}%
In addition, we know that
\begin{equation}
q^{2}=q_{3}^{2}-q_{0}^{2}\text{ \ and \ }p^{\prime 2}=-M^{2},
\end{equation}%
which lead us to
\begin{equation}
\omega ^{\prime }=\frac{2Mq_{0}-q^{2}}{2q_{0}+2M-2q_{3}\cos \theta }.
\end{equation}%
We write the polarization vectors explicitly in the following way:%
\begin{eqnarray}
\varepsilon ^{\mu }\left( q,0\right) &=&\frac{1}{q}\left(
q_{3},0,0,q_{0}\right) , \\
\varepsilon ^{\mu }\left( q,+1\right) &=&\left( 0,1,0,0\right) , \\
\varepsilon ^{\mu }\left( q,-1\right) &=&\left( 0,0,1,0\right) , \\
\varepsilon ^{\prime \mu }\left( q,+1\right) &=&\left( 0,\cos \theta
,0,-\sin \theta \right) , \\
\varepsilon ^{\prime \mu }\left( q,-1\right) &=&\left( 0,0,1,0\right) ,
\end{eqnarray}%
where all the transverse polarization vectors are normalized to $1$.
Therefore, one can easily average over $\lambda $ and sum over $\lambda
^{\prime }$, which yields
\begin{eqnarray}
\frac{1}{2}\sum_{\lambda ,\lambda ^{\prime }}\left\vert \mathcal{M}%
\right\vert ^{2} &=&2\mathcal{Q}^{4}\frac{c_{i}^{2}c_{f}^{2}}{z_{0}^{4}}%
\mathcal{C}_{1\text{DVCS}}^{2}\sum_{\lambda ,\lambda ^{\prime }}\left\vert
\varepsilon \left( q,\lambda \right) \cdot \varepsilon ^{\prime }\left(
q^{\prime },\lambda ^{\prime }\right) \right\vert ^{2}  \nonumber \\
&+&8\mathcal{Q}^{4}\frac{c_{i}^{2}c_{f}^{2}}{z_{0}^{4}}\mathcal{T}_{1\text{%
DVCS}}^{s2}\sum_{\lambda ,\lambda ^{\prime }}\left\vert \varepsilon \left(
q,\lambda \right) \cdot p\varepsilon ^{\prime }\left( q^{\prime },\lambda
^{\prime }\right) \cdot p^{\prime }\right\vert ^{2}  \nonumber \\
&+&8\mathcal{Q}^{4}\frac{c_{i}^{2}c_{f}^{2}}{z_{0}^{4}}\mathcal{C}_{1\text{%
DVCS}}\mathcal{T}_{1\text{DVCS}}^{s}\sum_{\lambda ,\lambda ^{\prime
}}\varepsilon \left( q,\lambda \right) \cdot p\varepsilon ^{\prime }\left(
q^{\prime },\lambda ^{\prime }\right) \cdot p^{\prime }\varepsilon \left(
q,\lambda \right) \cdot \varepsilon ^{\prime }\left( q^{\prime },\lambda
^{\prime }\right) .
\end{eqnarray}%
With the above polarization vectors, we find
\begin{eqnarray}
&&\sum_{\lambda ,\lambda ^{\prime }}\left\vert \varepsilon \left( q,\lambda
\right) \cdot \varepsilon ^{\prime }\left( q^{\prime },\lambda ^{\prime
}\right) \right\vert ^{2}=1+\cos ^{2}\theta +\frac{q_{0}^{2}}{q^{2}}\sin
^{2}\theta , \\
&&\sum_{\lambda ,\lambda ^{\prime }}\left\vert \varepsilon \left( q,\lambda
\right) \cdot p\varepsilon ^{\prime }\left( q^{\prime },\lambda ^{\prime
}\right) \cdot p^{\prime }\right\vert ^{2}=M^{2}q_{3}^{2}\frac{q_{3}^{2}}{%
q^{2}}\sin ^{2}\theta , \\
&&\sum_{\lambda ,\lambda ^{\prime }}\left[ \varepsilon \left( q,\lambda
\right) \cdot p\right] \left[ \varepsilon ^{\prime }\left( q^{\prime
},\lambda ^{\prime }\right) \cdot p^{\prime }\right] \left[ \varepsilon
\left( q,\lambda \right) \cdot \varepsilon ^{\prime }\left( q^{\prime
},\lambda ^{\prime }\right) \right] =-Mq_{0}\frac{q_{3}^{2}}{q^{2}}\sin
^{2}\theta .
\end{eqnarray}%
Therefore, the averaged amplitude can be written as
\begin{eqnarray}
\frac{1}{2}\sum_{\lambda ,\lambda ^{\prime }}\left\vert \mathcal{M}%
\right\vert ^{2} &\simeq &2\mathcal{Q}^{4}\frac{c_{i}^{2}c_{f}^{2}}{z_{0}^{4}%
}\mathcal{C}_{1\text{DVCS}}^{2}\left[ 1+\cos ^{2}\theta +\frac{q_{0}^{2}}{%
q^{2}}\sin ^{2}\theta \right]  \nonumber \\
&&+8\mathcal{Q}^{4}\frac{c_{i}^{2}c_{f}^{2}}{z_{0}^{4}}\mathcal{C}_{1\text{%
DVCS}}^{2}\left[ \frac{4M^{2}q_{3}^{2}}{\left( M^{2}+(p+q)^{2}\right) ^{2}}+%
\frac{4Mq_{0}}{M^{2}+(p+q)^{2}}\right] \frac{q_{3}^{2}}{q^{2}}\sin
^{2}\theta .
\end{eqnarray}%
The phase space integral in this frame is
\begin{equation}
\int d\Pi _{2}=\frac{1}{4\pi }\frac{\omega ^{\prime 2}}{2Mq_{0}-q^{2}}\int
d\cos \theta .
\end{equation}%
Plugging everything into the cross-section formula, one finds
\begin{eqnarray}
\frac{d\sigma ^{\text{Un}}}{d\cos \theta } &=&\frac{1}{2q_{0}}\frac{1}{2M}%
\frac{1}{4\pi }\frac{\omega ^{\prime 2}}{2Mq_{0}-q^{2}}\frac{1}{2}%
\sum_{\lambda ,\lambda ^{\prime }}\left\vert \mathcal{M}\right\vert ^{2} \\
&=&\frac{\omega ^{\prime 2}}{16M^{2}q_{0}^{2}\left( 1-x\right) }\mathcal{Q}%
^{4}\frac{c_{i}^{2}c_{f}^{2}}{z_{0}^{4}}\mathcal{C}_{1\text{DVCS}}^{2}\left[
1+\cos ^{2}\theta +\chi \sin ^{2}\theta \right] ,
\end{eqnarray}%
with
\begin{eqnarray}
\chi &=&\frac{q_{0}^{2}}{q^{2}}+\frac{q_{3}^{2}}{q^{2}}\left[ \frac{1}{%
\left( 1-x\right) ^{2}}\frac{q_{3}^{2}}{q_{0}^{2}}-\frac{2}{1-x}\right] \\
&=&\frac{q^{2}}{4M^{2}\left( 1-x\right) ^{2}}\left( 1+\frac{4M^{2}}{q^{2}}%
x\right) ^{2}.
\end{eqnarray}%
When $q^{2}\gg 4M^{2}$, one finds $\chi \gg 1$.

\section{Compton scattering Of dilatino}

\label{fermion}

In this section, we formulate the Compton scattering on a fermion target by
evaluating the \textit{s}-channel and \textit{u}-channel supergravity diagrams.

\subsection{Fermion field in AdS space and bulk to bulk propagator}

Spin-$\frac{1}{2}$ hadrons correspond to supergravity modes of the dilatino
which satisfy the Dirac equation in $\textrm{AdS}_{5}$ space. In the conformal region
the dilatino field can be written as
\begin{equation}
\lambda =\Psi (y,z)\otimes \eta (\Omega )\ ,
\end{equation}%
where $\Psi (y,z)$ is an $SO(4,1)$ spinor on $\textrm{AdS}_{5}$ and $\eta (\Omega )$
is an $SO(5)$ spinor on $S^{5}$. The wave function $\psi $ satisfies a
five-dimensional Dirac equation in AdS space. Let us first derive the five-dimensional
Dirac equation in the following.

Since the Dirac gamma matrices are defined in the flat spacetime, we need to
use vielbeins to convert them to curved spacetime and make the product $D\hspace{-8pt}\slash$ Lorentz invariant.
A convenient choice of such vielbeins is given by
\begin{equation}
e_{m}^{a}=\frac{R}{z}\delta _{m}^{a},\ \ e^{ma}=\frac{z}{R}\eta ^{ma}\ \
\text{and }\ e_{a}^{m}=\frac{z}{R}\delta _{a}^{m},
\end{equation}%
where $m=0,1,2,3,5$. The Levi-Civita connection, which is also known as the
Christoffel connection, is given by
\begin{equation}
\Gamma _{mn}^{p}=\frac{1}{2}g^{pq}(\partial _{n}g_{mq}+\partial
_{m}g_{nq}-\partial _{q}g_{mn})
\end{equation}%
Here we use $a,b,c$ to denote indices in flat space, and $m,n,p,q$ to denote
indices in curved space ($\textrm{AdS}_{5}$ space). In addition, the Greek indices $%
\mu ,\nu $ are defined in Minkowski space. From the metric, one knows
\begin{equation}
g_{mn}=\frac{R^{2}}{z^{2}}\eta _{mn}.
\end{equation}%
It is then straightforward to work out the Levi-Civita connection in $%
\textrm{AdS}_{5} $ space
\begin{equation}
\Gamma _{\mu \nu }^{5}=\frac{1}{z}\eta _{\mu \nu },\ \ \Gamma _{55}^{5}=-%
\frac{1}{z}\ \ \text{and }\ \Gamma _{\nu 5}^{\mu }=-\frac{1}{z}\delta _{\nu
}^{\mu }.
\end{equation}%
From the above vielbeins and Levi-Civita connections, we can compute the
spin connection which is defined as
\begin{equation}
\omega _{m}^{ab}=e_{n}^{a}\partial _{m}e^{nb}+e_{n}^{a}e^{pb}\Gamma
_{pm}^{n}.
\end{equation}%
The only nonvanishing spin connections are%
\begin{equation}
\omega _{\mu }^{5\nu }=-\omega _{\mu }^{\nu 5}=\frac{1}{z}\delta _{\mu
}^{\nu }.
\end{equation}%
Using the above results, the operator $D\hspace{-8pt}\slash$ can be cast into
\begin{equation}
D\hspace{-8pt}\slash=g^{mn}e_{n}^{a}\gamma _{a}\left( \partial _{m}+\frac{1}{%
2}\omega _{m}^{bc}\Sigma _{bc}\right) =\frac{z}{R}\left( \gamma ^{5}\partial
_{z}+\gamma ^{\mu }\partial _{\mu }-\frac{2}{z}\gamma ^{5}\right) ,
\end{equation}%
with $\Sigma _{\mu 5}=\frac{1}{4}\left[ \gamma _{\mu },\gamma _{5}\right]$.%
\footnote{%
We define the $\gamma$-matrices according to the Dirac algebra in the mostly
plus metric signature $-+++$. The $\gamma _{\mu }$ that we used differ from
the conventional gamma matrices by a factor of $i$.} The free dilatino field
in $\textrm{AdS}_{5}$ space satisfies the Dirac equation \cite{Henningson:1998cd,
Muck:1998rr}
\begin{equation}
(D\hspace{-8pt}\slash-m)\Psi =\frac{z}{R}\left( \gamma ^{5}\partial
_{z}+\gamma ^{\mu }\partial _{\mu }-\frac{2}{z}\gamma ^{5}-\frac{mR}{z}%
\right) \Psi =0.
\end{equation}%
Its normalizable solution is given by
\begin{equation}
\Psi (z,y)=Ce^{ip\cdot y}z^{\frac{5}{2}}\left[
J_{mR-1/2}(Mz)P_{+}+J_{mR+1/2}(Mz)P_{-}\right] u_{\sigma }
\end{equation}%
with
\begin{equation}
p\hspace{-4.5pt}\slash u_{\sigma }=-iMu_{\sigma }(\sigma =1,2),\ \
M^{2}=-p^{2},\ \ P_{\pm }=\frac{1}{2}(1\pm \gamma ^{5}).
\end{equation}%
Therefore, we choose the initial and final wave functions of the dilatino to
be
\begin{eqnarray}
\Psi _{i}(y,z) &=&\frac{c_{i}^{\prime }}{z_{0}R^{9/2}}e^{ip\cdot y}z^{5/2}%
\left[ J_{mR-1/2}(Mz)P_{+}+J_{mR+1/2}(Mz)P_{-}\right] u_{i\sigma }  \nonumber
\\
\overline{\Psi }_{f}^{\ast }(y^{\prime },z^{\prime }) &=&\frac{c_{f}^{\prime
}}{z_{0}R^{9/2}}e^{-ip^{\prime }\cdot y^{\prime }}\overline{u}_{f\sigma
^{\prime }}z^{\prime 5/2}\left[ J_{mR-1/2}(Mz^{\prime
})P_{-}+J_{mR+1/2}(Mz^{\prime })P_{+}\right] ,
\end{eqnarray}%
with a chosen boundary condition $J_{mR-1/2}(Mz_{0})=0$.

Furthermore, we can now consider the bulk to bulk propagator of dilatino. It
satisfies the following differential equation
\begin{equation}
\frac{z}{R}\left( \gamma ^{5}\partial _{z}+\gamma ^{\mu }\partial _{\mu }-%
\frac{2}{z}\gamma ^{5}-\frac{mR}{z}\right) G(z,y;z^{\prime },y^{\prime })=%
\frac{z^{5}}{R}\delta (z-z^{\prime })\delta ^{4}(y-y^{\prime }).
\end{equation}%
Using the identity
\begin{eqnarray}
&&(D\hspace{-8pt}\slash-m)\left\{ e^{ik\cdot y}z^{\frac{5}{2}}\left[
J_{mR-1/2}(\omega z)P_{+}+J_{mR+1/2}(\omega z)P_{-}\right] \right\}
\nonumber \\
&=&\frac{z}{R}\left\{ e^{ip\cdot y}z^{\frac{5}{2}}\left[ J_{mR-1/2}(\omega
z)P_{-}+J_{mR+1/2}(\omega z)P_{+}\right] \right\} \left( ik\hspace{-6pt}%
\slash-\omega \right) ,  \label{greenfunction}
\end{eqnarray}%
it is to straightforward to verify that the solution of Eq.~(\ref%
{greenfunction}) is given by
\begin{eqnarray}
G(y,z;y^{\prime },z^{\prime }) &=&-\int \frac{d^{4}k}{(2\pi )^{4}}%
e^{-ik\cdot (y-y^{\prime })}\int_{0}^{\infty }d\omega \omega z^{5/2}\left[
J_{mR-1/2}(\omega z)P_{+}+J_{mR+1/2}(\omega z)P_{-}\right]  \nonumber \\
&&\times \frac{-ik\hspace{-6pt}\slash+\omega }{\omega ^{2}+k^{2}-i\epsilon }%
z^{\prime 5/2}\left[ J_{mR-1/2}(\omega z^{\prime })P_{-}+J_{mR+1/2}(\omega
z^{\prime })P_{+}\right] .
\end{eqnarray}%
Following the usual convention, we start with the final state fermion and
end with the initial state fermion when we write the amplitude. Thus the
following Green's function is used hereafter:
\begin{eqnarray}
G(z^{\prime },y^{\prime };z,y) &=&-\int \frac{d^{4}k}{(2\pi )^{4}}e^{ik\cdot
(y^{\prime }-y)}\int_{0}^{\infty }d\omega \omega z^{\prime }{}^{5/2}\left[
J_{mR-1/2}(\omega z^{\prime })P_{+}+J_{mR+1/2}(\omega z^{\prime })P_{-}%
\right]  \nonumber \\
&&\times \frac{ik\hspace{-6pt}\slash+\omega }{\omega ^{2}+k^{2}-i\epsilon }%
z^{5/2}\left[ J_{mR-1/2}(\omega z)P_{-}+J_{mR+1/2}(\omega z)P_{+}\right] .
\end{eqnarray}

\subsection{Compton scattering amplitudes}

The interaction vertex between the Kaluza-Klein gauge field and dilatino is
given by%
\begin{eqnarray}
S_{int} &=&\int d^{10}x\sqrt{-g}\,A_{m}v^{w}\bar{\lambda}_{X}\gamma
^{m}\partial _{w}\lambda _{{i}} \\
&=&i{\mathcal{Q}}R^{5}\int d^{5}x\sqrt{-g}\,A_{m}\overline{\Psi }_{X}^{\ast }%
{e^{m}}_{a}\gamma ^{a}\Psi _{{i}},
\end{eqnarray}%
where $\lambda $ stands for the full 10-dimensional wave function of
the dilatino state. In arriving at the above expression, we have used the fact that $%
v^{w}\partial _{w}\lambda _{{i}}=i{\mathcal{Q}}\lambda _{{i}}$
and the integration over $S^{5}$ gives $R^{5}$.

Now we are ready to calculate the Compton scattering amplitude of the dilatino
from gauge/string duality. In this section, we consider the following two
diagrams, namely, the \textit{s}-channel and \textit{u}-channel graphs. The \textit{s}-channel is
illustrated in Fig.~\ref{DVCS}. Here we use the bulk to bulk propagator from
$z^{\prime }$ to $z$, which is different from the deeply inelastic
scattering where the propagator is always taken to be on shell. Using the
dilatino bulk to bulk propagator and above interaction vertex, we can write
\textit{s}-channel amplitude $T_{\mu \nu }^{s}$ as
\[
T_{\mu \nu }^{s}=T_{\mu \nu 1}^{s}+T_{\mu \nu 2}^{s}+T_{\mu \nu
3}^{s}+T_{\mu \nu 4}^{s},
\]%
where
\begin{eqnarray}
T_{1\mu \nu }^{s} &=&\mathcal{A}^{\prime }\int dz\int dz^{\prime
}z^{-1/2}z^{\prime }{}^{-1/2}q^{\prime }K_{1}(q^{\prime }z^{\prime
})qK_{1}(qz)  \nonumber  \label{ts1} \\
&&\times \bar{u}_{\sigma ^{\prime }}(p^{\prime })\left[ J_{mR-1/2}(Mz^{%
\prime })P_{-}+J_{mR+1/2}(Mz^{\prime })P_{+}\right] \gamma _{\mu }  \nonumber
\\
&&\times G(z^{\prime },z;p+q)\gamma _{\nu }\left[
J_{mR-1/2}(Mz)P_{+}+J_{mR+1/2}(Mz)P_{-}\right] u_{\sigma }(p),
\end{eqnarray}%
\begin{eqnarray}
T_{2\mu \nu }^{s} &=&i\mathcal{A}^{\prime }\int dz\int dz^{\prime
}z^{-1/2}z^{\prime }{}^{-1/2}q^{\prime }K_{1}(q^{\prime }z^{\prime })q_{\nu
}K_{0}(qz)  \nonumber \\
&&\times \bar{u}_{\sigma ^{\prime }}(p^{\prime })\left[ J_{mR-1/2}(Mz^{%
\prime })P_{-}+J_{mR+1/2}(Mz^{\prime })P_{+}\right] \gamma _{\mu }  \nonumber
\\
&&\times G(z^{\prime },z;p+q)\gamma ^{5}\left[
J_{mR-1/2}(Mz)P_{+}+J_{mR+1/2}(Mz)P_{-}\right] u_{\sigma }(p),
\end{eqnarray}%
\begin{eqnarray}
T_{3\mu \nu }^{s} &=&-i\mathcal{A}^{\prime }\int dz\int dz^{\prime
}z^{-1/2}z^{\prime }{}^{-1/2}q_{\mu }^{\prime }K_{0}(q^{\prime }z^{\prime
})qK_{1}(qz)  \nonumber \\
&&\times \bar{u}_{\sigma ^{\prime }}(p^{\prime })\left[ J_{mR-1/2}(Mz^{%
\prime })P_{-}+J_{mR+1/2}(Mz^{\prime })P_{+}\right] \gamma ^{5}  \nonumber \\
&&\times G(z^{\prime },z;p+q)\gamma _{\nu }\left[
J_{mR-1/2}(Mz)P_{+}+J_{mR+1/2}(Mz)P_{-}\right] u_{\sigma }(p),
\end{eqnarray}%
\begin{eqnarray}
T_{4\mu \nu }^{s} &=&\mathcal{A}^{\prime }\int dz\int dz^{\prime
}z^{-1/2}z^{\prime }{}^{-1/2}q_{\mu }^{\prime }K_{0}(q^{\prime }z^{\prime
})q_{\nu }K_{0}(qz)  \nonumber \\
&&\times \bar{u}_{\sigma ^{\prime }}(p^{\prime })\left[ J_{mR-1/2}(Mz^{%
\prime })P_{-}+J_{mR+1/2}(Mz^{\prime })P_{+}\right] \gamma ^{5}  \nonumber \\
&&\times G(z^{\prime },z;p+q)\gamma ^{5}\left[
J_{mR-1/2}(Mz)P_{+}+J_{mR+1/2}(Mz)P_{-}\right] u_{\sigma }(p),
\end{eqnarray}%
with $\mathcal{A}^{\prime }=\frac{\mathcal{Q}^{2}c_{i}^{\prime
}c_{f}^{\prime }}{z_{0}^{2}}(2\pi )^{4}\delta ^{(4)}(p+q-p^{\prime
}-q^{\prime })$\footnote{%
Here we assign each dilatino propagator a factor of $\frac{1}{R^{9}}$, and
find that the Compton amplitude is independent of $R$.} and
\begin{eqnarray}
G(z^{\prime },z;k) &=&-\int_{0}^{\infty }d\omega \omega z^{\prime }{}^{5/2}
\left[ J_{mR-1/2}(\omega z^{\prime })P_{+}+J_{mR+1/2}(\omega z^{\prime
})P_{-}\right]  \nonumber \\
&&\times \frac{ik\hspace{-6pt}\slash+\omega }{\omega ^{2}+k^{2}-i\epsilon }%
z^{5/2}\left[ J_{mR-1/2}(\omega z)P_{-}+J_{mR+1/2}(\omega z)P_{+}\right] .
\end{eqnarray}

Using Eq.~(\ref{kr}) and $p\hspace{-4.5pt}\slash u_{\sigma }=-iMu_{\sigma }$%
, along with the recurrence relations
\begin{eqnarray}
\frac{d}{dx}J_{n}(x) &=&J_{n-1}(x)-\frac{n}{x}J_{n}(x), \\
\frac{d}{dx}J_{n}(x) &=&\frac{n}{x}J_{n}(x)-J_{n+1}(x),
\end{eqnarray}%
we can have
\begin{eqnarray}
&&\int dzz^{2}K_{0}(qz)(ik\hspace{-6pt}\slash+\omega )\left[
J_{mR-1/2}(\omega z)P_{-}+J_{mR+1/2}(\omega z)P_{+}\right]  \nonumber \\
&&\times \gamma ^{5}\left[ J_{mR-1/2}(Mz)P_{+}+J_{mR+1/2}(Mz)P_{-}\right]
u_{\sigma }(p)  \nonumber \\
&=&\int dzz^{2}qK_{1}(qz)(ik\hspace{-6pt}\slash+\omega )\left[
J_{mR-1/2}(\omega z)P_{-}+J_{mR+1/2}(\omega z)P_{+}\right] \frac{iq\hspace{%
-6pt}\slash}{q^{2}}  \nonumber \\
&&\times \left[ J_{mR-1/2}(Mz)P_{+}+J_{mR+1/2}(Mz)P_{-}\right] u_{\sigma }(p)
\nonumber \\
&&+\int dzz^{2}qK_{1}(qz)\frac{\omega ^{2}+k^{2}}{q^{2}}\left[
J_{mR-1/2}(\omega z)P_{+}+J_{mR+1/2}(\omega z)P_{-}\right]  \nonumber \\
&&\times \left[ J_{mR-1/2}(Mz)P_{+}+J_{mR+1/2}(Mz)P_{-}\right] u_{\sigma
}(p).
\end{eqnarray}%
Similarly, we can also find
\begin{eqnarray}
&&\int dzz^{2}K_{0}(q^{\prime }z)\bar{u}_{\sigma ^{\prime }}(p^{\prime })%
\left[ J_{mR-1/2}(Mz)P_{-}+J_{mR+1/2}(Mz)P_{+}\right]  \nonumber \\
&&\times \gamma ^{5}\left[ J_{mR-1/2}(\omega z)P_{+}+J_{mR+1/2}(\omega
z)P_{-}\right] (ik\hspace{-6pt}\slash+\omega )  \nonumber \\
&=&-\int dzz^{2}q^{\prime }K_{1}(q^{\prime }z)\bar{u}_{\sigma ^{\prime
}}(p^{\prime })\left[ J_{mR-1/2}(Mz)P_{-}+J_{mR+1/2}(Mz)P_{+}\right]
\nonumber \\
&&\times \frac{iq^{\prime }\hspace{-8pt}\slash}{q^{\prime }{}^{2}}\left[
J_{mR-1/2}(\omega z)P_{+}+J_{mR+1/2}(\omega z)P_{-}\right] (ik\hspace{-6pt}%
\slash+\omega )  \nonumber \\
&&-\int dzz^{2}q^{\prime }K_{1}(q^{\prime }z)\bar{u}_{\sigma ^{\prime
}}(p^{\prime })\left[ J_{mR-1/2}(Mz)P_{-}+J_{mR+1/2}(Mz)P_{+}\right]
\nonumber \\
&&\times \left[ J_{mR-1/2}(\omega z)P_{-}+J_{mR+1/2}(\omega z)P_{+}\right]
\frac{\omega ^{2}+k^{2}}{q^{\prime }{}^{2}}.
\end{eqnarray}

This can simplify the Compton scattering amplitude a lot, and lead
to
\begin{eqnarray}
T_{2\mu \nu }^{s} &=&-\mathcal{A}^{\prime }\int dz\int dz^{\prime
}z^{-1/2}z^{\prime }{}^{-1/2}q^{\prime }K_{1}(q^{\prime }z^{\prime
})qK_{1}(qz)  \nonumber \\
&&\times \bar{u}_{\sigma ^{\prime }}(p^{\prime })\left[ J_{mR-1/2}(Mz^{%
\prime })P_{-}+J_{mR+1/2}(Mz^{\prime })P_{+}\right] \gamma _{\mu }  \nonumber
\\
&&\times G(z^{\prime },z;p+q)\frac{q_{\nu }q\hspace{-5pt}\slash}{q^{2}}\left[
J_{mR-1/2}(Mz)P_{+}+J_{mR+1/2}(Mz)P_{-}\right] u_{\sigma }(p)  \nonumber \\
&&-i\mathcal{A}^{\prime }\bar{u}_{\sigma ^{\prime }}(p^{\prime })\gamma
_{\mu }P_{+}u_{\sigma }(p)\frac{q_{\nu }}{q}\int
dzz^{3}K_{1}(qz)K_{1}(q^{\prime }z)J_{mR-1/2}^{2}(Mz)  \nonumber \\
&&-i\mathcal{A}^{\prime }\bar{u}_{\sigma ^{\prime }}(p^{\prime })\gamma
_{\mu }P_{-}u_{\sigma }(p)\frac{q_{\nu }}{q}\int
dzz^{3}K_{1}(qz)K_{1}(q^{\prime }z)J_{mR+1/2}^{2}(Mz), \\
T_{3\mu \nu }^{s} &=&-\mathcal{A}^{\prime }\int dz\int dz^{\prime
}z^{-1/2}z^{\prime }{}^{-1/2}q^{\prime }K_{1}(q^{\prime }z^{\prime
})qK_{1}(qz)  \nonumber \\
&&\times \bar{u}_{\sigma ^{\prime }}(p^{\prime })\left[ J_{mR-1/2}(Mz^{%
\prime })P_{-}+J_{mR+1/2}(Mz^{\prime })P_{+}\right] \frac{q_{\mu }^{\prime
}q^{\prime }\hspace{-7.5pt}\slash}{q^{\prime }{}^{2}}  \nonumber \\
&&\times G(z^{\prime },z;p+q)\gamma _{\nu }\left[
J_{mR-1/2}(Mz)P_{+}+J_{mR+1/2}(Mz)P_{-}\right] u_{\sigma }(p)  \nonumber \\
&&-i\mathcal{A}^{\prime }\bar{u}_{\sigma ^{\prime }}(p^{\prime })\gamma
_{\nu }P_{+}u_{\sigma }(p)\frac{q_{\mu }^{\prime }}{q^{\prime }}\int
dzz^{3}K_{1}(qz)K_{1}(q^{\prime }z)J_{mR-1/2}^{2}(Mz)  \nonumber \\
&&-i\mathcal{A}^{\prime }\bar{u}_{\sigma ^{\prime }}(p^{\prime })\gamma
_{\nu }P_{-}u_{\sigma }(p)\frac{q_{\mu }^{\prime }}{q^{\prime }}\int
dzz^{3}K_{1}(qz)K_{1}(q^{\prime }z)J_{mR+1/2}^{2}(Mz)
\end{eqnarray}%
and
\begin{eqnarray}
T_{4\mu \nu }^{s} &=&\mathcal{A}^{\prime }\int dz\int dz^{\prime
}z^{-1/2}z^{\prime }{}^{-1/2}q^{\prime }K_{1}(q^{\prime }z^{\prime
})qK_{1}(qz)  \nonumber \\
&&\times \bar{u}_{\sigma ^{\prime }}(p^{\prime })\left[ J_{mR-1/2}(Mz^{%
\prime })P_{-}+J_{mR+1/2}(Mz^{\prime })P_{+}\right] \frac{q_{\mu }^{\prime
}q^{\prime }\hspace{-7.5pt}\slash}{q^{\prime }{}^{2}}  \nonumber \\
&&\times G(z^{\prime },z;p+q)\frac{q_{\nu }q\hspace{-6.5pt}\slash}{q^{2}}%
\left[ J_{mR-1/2}(Mz)P_{+}+J_{mR+1/2}(Mz)P_{-}\right] u_{\sigma }(p)
\nonumber \\
&&+i\mathcal{A}^{\prime }\bar{u}_{\sigma ^{\prime }}(p^{\prime })q\hspace{%
-6.5pt}\slash P_{+}u_{\sigma }(p)\frac{q_{\mu }^{\prime }q_{\nu }}{q^{\prime
}q}\int dzz^{3}K_{1}(q^{\prime }z)K_{1}(qz)J_{mR-1/2}^{2}(Mz)  \nonumber \\
&&+i\mathcal{A}^{\prime }\bar{u}_{\sigma ^{\prime }}(p^{\prime })q\hspace{%
-6.5pt}\slash P_{-}u_{\sigma }(p)\frac{q_{\mu }^{\prime }q_{\nu }}{q^{\prime
}q}\int dzz^{3}K_{1}(q^{\prime }z)K_{1}(qz)J_{mR+1/2}^{2}(Mz)  \nonumber \\
&&-\mathcal{A}^{\prime }\bar{u}_{\sigma ^{\prime }}(p^{\prime })\gamma
^{5}u_{\sigma }(p)\frac{q_{\mu }^{\prime }q_{\nu }}{q^{\prime }q}\int
dzz^{3}q^{\prime }K_{0}(q^{\prime }z)K_{1}(qz)J_{mR-1/2}(Mz)J_{mR+1/2}(Mz).
\end{eqnarray}

Using the same trick, it is then straightforward to calculate the \textit{u}-channel
graph as illustrated in Fig.~(\ref{DVCS2}) and obtain
\begin{equation}
T_{\mu \nu }^{u}=T_{1\mu \nu }^{u}+T_{2\mu \nu }^{u}+T_{3\mu \nu
}^{u}+T_{4\mu \nu }^{u},
\end{equation}%
where
\begin{eqnarray}
T_{1\mu \nu }^{u} &=&\mathcal{A}^{\prime }\int dz\int dz^{\prime
}z^{-1/2}z^{\prime }{}^{-1/2}qK_{1}(qz^{\prime })q^{\prime }K_{1}(q^{\prime
}z)  \nonumber \\
&&\times \bar{u}_{\sigma ^{\prime }}(p^{\prime })\left[ J_{mR-1/2}(Mz^{%
\prime })P_{-}+J_{mR+1/2}(Mz^{\prime })P_{+}\right] \gamma _{\nu }  \nonumber
\\
&&\times G(z^{\prime },z;p-q^{\prime })\gamma _{\mu }\left[
J_{mR-1/2}(Mz)P_{+}+J_{mR+1/2}(Mz)P_{-}\right] u_{\sigma }(p),  \label{tu1}
\end{eqnarray}

\begin{eqnarray}
T_{2\mu \nu }^{u} &=&-\mathcal{A}^{\prime }\int dz\int dz^{\prime
}z^{-1/2}z^{\prime }{}^{-1/2}qK_{1}(qz^{\prime })q^{\prime }K_{1}(q^{\prime
}z)  \nonumber \\
&&\times \bar{u}_{\sigma ^{\prime }}(p^{\prime })\left[ J_{mR-1/2}(Mz^{%
\prime })P_{-}+J_{mR+1/2}(Mz^{\prime })P_{+}\right] \gamma _{\nu }  \nonumber
\\
&&\times G(z^{\prime },z;p-q^{\prime })\frac{q_{\mu }^{\prime }q^{\prime }%
\hspace{-7.5pt}\slash}{q^{\prime }{}^{2}}\left[
J_{mR-1/2}(Mz)P_{+}+J_{mR+1/2}(Mz)P_{-}\right] u_{\sigma }(p)  \nonumber \\
&&+i\mathcal{A}^{\prime }\bar{u}_{\sigma ^{\prime }}(p^{\prime })\gamma
_{\nu }P_{+}u_{\sigma }(p)\frac{q_{\mu }^{\prime }}{q^{\prime }}\int
dzz^{3}K_{1}(qz)K_{1}(q^{\prime }z)J_{mR-1/2}^{2}(Mz)  \nonumber \\
&&+i\mathcal{A}^{\prime }\bar{u}_{\sigma ^{\prime }}(p^{\prime })\gamma
_{\nu }P_{-}u_{\sigma }(p)\frac{q_{\mu }^{\prime }}{q^{\prime }}\int
dzz^{3}K_{1}(qz)K_{1}(q^{\prime }z)J_{mR+1/2}^{2}(Mz),  \label{tu2} \\
T_{3\mu \nu }^{u} &=&-\mathcal{A}^{\prime }\int dz\int dz^{\prime
}z^{-1/2}z^{\prime }{}^{-1/2}qK_{1}(qz^{\prime })q^{\prime }K_{1}(q^{\prime
}z)  \nonumber \\
&&\times \bar{u}_{\sigma ^{\prime }}(p^{\prime })\left[ J_{mR-1/2}(Mz^{%
\prime })P_{-}+J_{mR+1/2}(Mz^{\prime })P_{+}\right] \frac{q_{\nu }q\hspace{%
-7pt}\slash}{q^{2}}  \nonumber \\
&&\times G(z^{\prime },z;p-q^{\prime })\gamma _{\mu }\left[
J_{mR-1/2}(Mz)P_{+}+J_{mR+1/2}(Mz)P_{-}\right] u_{\sigma }(p)  \nonumber \\
&&+i\mathcal{A}^{\prime }\bar{u}_{\sigma ^{\prime }}(p^{\prime })\gamma
_{\mu }P_{+}u_{\sigma }(p)\frac{q_{\nu }}{q}\int
dzz^{3}K_{1}(qz)K_{1}(q^{\prime }z)J_{mR-1/2}^{2}(Mz)  \nonumber \\
&&+i\mathcal{A}^{\prime }\bar{u}_{\sigma ^{\prime }}(p^{\prime })\gamma
_{\mu }P_{-}u_{\sigma }(p)\frac{q_{\nu }}{q}\int
dzz^{3}K_{1}(qz)K_{1}(q^{\prime }z)J_{mR+1/2}^{2}(Mz)  \label{tu3}
\end{eqnarray}%
and
\begin{eqnarray}
T_{4\mu \nu }^{u} &=&\mathcal{A}^{\prime }\int dz\int dz^{\prime
}z^{-1/2}z^{\prime }{}^{-1/2}qK_{1}(qz^{\prime })q^{\prime }K_{1}(q^{\prime
}z)  \nonumber  \label{tu4} \\
&&\times \bar{u}_{\sigma ^{\prime }}(p^{\prime })\left[ J_{mR-1/2}(Mz^{%
\prime })P_{-}+J_{mR+1/2}(Mz^{\prime })P_{+}\right] \frac{q_{\nu }q\hspace{%
-6.5pt}\slash}{q^{2}}  \nonumber \\
&&\times G(z^{\prime },z;p-q^{\prime })\frac{q_{\mu }^{\prime }q^{\prime }%
\hspace{-7.5pt}\slash}{q^{\prime }{}^{2}}\left[
J_{mR-1/2}(Mz)P_{+}+J_{mR+1/2}(Mz)P_{-}\right] u_{\sigma }(p)  \nonumber \\
&&-i\mathcal{A}^{\prime }\bar{u}_{\sigma ^{\prime }}(p^{\prime })q^{\prime }%
\hspace{-7.5pt}\slash P_{+}u_{\sigma }(p)\frac{q_{\mu }^{\prime }q_{\nu }}{%
q^{\prime }q}\int dzz^{3}K_{1}(q^{\prime }z)K_{1}(qz)J_{mR-1/2}^{2}(Mz)
\nonumber \\
&&-i\mathcal{A}^{\prime }\bar{u}_{\sigma ^{\prime }}(p^{\prime })q^{\prime }%
\hspace{-7.5pt}\slash P_{-}u_{\sigma }(p)\frac{q_{\mu }^{\prime }q_{\nu }}{%
q^{\prime }q}\int dzz^{3}K_{1}(q^{\prime }z)K_{1}(qz)J_{mR+1/2}^{2}(Mz)
\nonumber \\
&&-\mathcal{A}^{\prime }\bar{u}_{\sigma ^{\prime }}(p^{\prime })\gamma
^{5}u_{\sigma }(p)\frac{q_{\mu }^{\prime }q_{\nu }}{q^{\prime }}\int
dzz^{3}K_{0}(qz)K_{1}(q^{\prime }z)J_{mR-1/2}(Mz)J_{mR+1/2}(Mz)
\end{eqnarray}

After some more algebra, one can find the total contribution is given by
\begin{eqnarray}
T_{\mu \nu } &=&T_{\mu \nu }^{s}+T_{\mu \nu }^{u}  \nonumber  \label{total1}
\\
&=&\mathcal{A}^{\prime }\int dz\int dz^{\prime }z^{-1/2}z^{\prime
}{}^{-1/2}q^{\prime }K_{1}(q^{\prime }z^{\prime })qK_{1}(qz)  \nonumber \\
&&\times \bar{u}_{\sigma ^{\prime }}(p^{\prime })\left[ J_{mR-1/2}(Mz^{%
\prime })P_{-}+J_{mR+1/2}(Mz^{\prime })P_{+}\right] \left( \gamma _{\mu }-%
\frac{q_{\mu }^{\prime }q^{\prime }\hspace{-7.5pt}\slash}{q^{\prime }{}^{2}}%
\right)  \nonumber \\
&&\times G(z^{\prime },z;p+q)\left( \gamma _{\nu }-\frac{q_{\nu }q\hspace{%
-6.5pt}\slash}{q^{2}}\right) \left[ J_{mR-1/2}(Mz)P_{+}+J_{mR+1/2}(Mz)P_{-}%
\right] u_{\sigma }(p)  \nonumber \\
&&+\mathcal{A}^{\prime }\int dz\int dz^{\prime }z^{-1/2}z^{\prime
}{}^{-1/2}qK_{1}(qz^{\prime })q^{\prime }K_{1}(q^{\prime }z)  \nonumber \\
&&\times \bar{u}_{\sigma ^{\prime }}(p^{\prime })\left[ J_{mR-1/2}(Mz^{%
\prime })P_{-}+J_{mR+1/2}(Mz^{\prime })P_{+}\right] \left( \gamma _{\nu }-%
\frac{q_{\nu }q\hspace{-6.5pt}\slash}{q^{2}}\right)  \nonumber \\
&&\times G(z^{\prime },z;p-q^{\prime })\left( \gamma _{\mu }-\frac{q_{\mu
}^{\prime }q^{\prime }\hspace{-7.5pt}\slash}{q^{\prime }{}^{2}}\right) \left[
J_{mR-1/2}(Mz)P_{+}+J_{mR+1/2}(Mz)P_{-}\right] u_{\sigma }(p).
\end{eqnarray}%
It is obvious that the current conservation, namely, $T_{\mu \nu }q^{\nu }=0$
and $q^{\prime \mu }T_{\mu \nu }=0$, is satisfied as a result of
cancellation between \textit{s}-channel and \textit{u}-channel amplitudes due to gauge
invariance. In addition, Eq.~(\ref{total1}) can be recast into a simple form
\begin{eqnarray}
T_{\mu \nu } &=&-\mathcal{A}^{\prime }e_{\mu \alpha }^{L}(q^{\prime })e_{\nu
\beta }^{L}(q)\int_{0}^{\infty }d\omega \frac{\omega }{\omega
^{2}+(p+q)^{2}-i\epsilon }\bar{u}_{\sigma ^{\prime }}(p^{\prime })\left[
\mathcal{C}_{1}(q^{\prime },\omega )P_{-}+\mathcal{C}_{2}(q^{\prime },\omega
)P_{+}\right]  \nonumber  \label{total2} \\
&&\times \gamma ^{\alpha }\left[ i(p\hspace{-6pt}\slash+q\hspace{-6pt}\slash%
)+\omega \right] \gamma ^{\beta }\left[ \mathcal{C}_{1}(q,\omega )P_{+}+%
\mathcal{C}_{2}(q,\omega )P_{-}\right] u_{\sigma }(p)  \nonumber \\
&&-\mathcal{A}^{\prime }e_{\mu \beta }^{L}(q^{\prime })e_{\nu \alpha
}^{L}(q)\int_{0}^{\infty }d\omega \frac{\omega }{\omega ^{2}+(p-q^{\prime
})^{2}-i\epsilon }\bar{u}_{\sigma ^{\prime }}(p^{\prime })\left[ \mathcal{C}%
_{1}(q,\omega )P_{-}+\mathcal{C}_{2}(q,\omega )P_{+}\right]  \nonumber \\
&&\times \gamma ^{\alpha }\left[ i(p\hspace{-6pt}\slash-q^{\prime }\hspace{%
-7pt}\slash)+\omega \right] \gamma ^{\beta }\left[ \mathcal{C}_{1}(q^{\prime
},\omega )P_{+}+\mathcal{C}_{2}(q^{\prime },\omega )P_{-}\right] u_{\sigma
}(p),
\end{eqnarray}%
where
\begin{eqnarray}
e_{\mu \alpha }^{L}(q) &=&g_{\mu \alpha }-\frac{q_{\mu }q_{\alpha }}{q^{2}}
\\
\mathcal{C}_{1}(q,\omega )
&=&\int_{0}^{z_{0}}dzz^{2}qK_{1}(qz)J_{mR-1/2}(Mz)J_{mR-1/2}(\omega z) \\
\mathcal{C}_{2}(q,\omega )
&=&\int_{0}^{z_{0}}dzz^{2}qK_{1}(qz)J_{mR+1/2}(Mz)J_{mR+1/2}(\omega z).
\end{eqnarray}

\subsection{Real Compton Scattering}

In the case of real Compton scattering, both initial and final state photons
are real. In the limit $q^{2}=0$, choosing the boundary condition $J_{\tau
-2}(Mz_{0})=0$ with $\tau =mR+3/2$, we can have

\begin{eqnarray}
\mathcal{C}_{1}(0,\omega ) &=&\frac{M}{M^{2}-\omega ^{2}}z_{0}J_{\tau
-2}(\omega z_{0})J_{\tau -1}(Mz_{0}) \\
\mathcal{C}_{2}(0,\omega ) &=&\frac{\omega }{M^{2}-\omega ^{2}}z_{0}J_{\tau
-2}(\omega z_{0})J_{\tau -1}(Mz_{0})
\end{eqnarray}%
Here we would like to work in a limit that $M\rightarrow 0$ and $%
z_{0}\rightarrow \infty $ but with $Mz_{0}$ fixed. If we take the limit $%
z_{0}\rightarrow \infty $, we can obtain
\begin{equation}
\mathcal{C}_{1}(0,\omega )\rightarrow \frac{1}{M}\delta (\omega -M)\text{
and }\mathcal{C}_{2}(0,\omega )\rightarrow \frac{1}{M}\delta (\omega -M).
\end{equation}%
Setting both $q^{2}=0$ and $q^{\prime 2}=0$, the real Compton scattering
amplitude is found to be
\begin{eqnarray}
T_{\mu \nu } &=&-\mathcal{A}^{\prime }z_{0}^{2}J_{\tau -1}^{2}(Mz_{0})\bar{u}%
_{\sigma ^{\prime }}(p^{\prime })\gamma _{\mu }\frac{i(p\hspace{-6pt}\slash+q%
\hspace{-6pt}\slash)+M}{M^{2}+(p+q)^{2}}\gamma _{\nu }u_{\sigma }(p)
\nonumber  \label{trcs2} \\
&&-\mathcal{A}^{\prime }z_{0}^{2}J_{\tau -1}^{2}(Mz_{0})\bar{u}_{\sigma
^{\prime }}(p^{\prime })\gamma _{\nu }\frac{i(p\hspace{-6pt}\slash-q^{\prime
}\hspace{-7pt}\slash)+M}{M^{2}+(p-q^{\prime })^{2}}\gamma _{\mu }u_{\sigma
}(p).
\end{eqnarray}%
In such approximation, the real Compton scattering amplitude found above is
parametrically the same as the Compton scattering amplitude of the
fundamental particle of spin $1/2$.

\subsection{Deeply virtual Compton scattering}

In the case of DVCS, the incoming photon is virtual, and the outgoing photon
is assumed to be real. Therefore, we set $q^{\prime 2}=0$ while $q^{2}\gg
M^{2}$, and still let $z_{0}\rightarrow \infty $ but with $Mz_{0}$ kept
fixed. Thus, the scattering amplitude in Eq.~(\ref{total2}) is found to be
\begin{eqnarray}
T_{\mu \nu } &=&-\mathcal{A}^{\prime }e_{\nu \beta }^{L}(q)\bar{u}_{\sigma
^{\prime }}(p^{\prime })\gamma _{\mu }\frac{i(p\hspace{-6pt}\slash+q\hspace{%
-6pt}\slash)+M}{M^{2}+(p+q)^{2}}\gamma ^{\beta }\left[ \mathcal{C}%
_{1}(q,M)P_{+}+\mathcal{C}_{2}(q,M)P_{-}\right] u_{\sigma }(p)  \nonumber
\label{trcs1} \\
&&-\mathcal{A}^{\prime }e_{\nu \alpha }^{L}(q)\bar{u}_{\sigma ^{\prime
}}(p^{\prime })\left[ \mathcal{C}_{1}(q,M)P_{-}+\mathcal{C}_{2}(q,M)P_{+}%
\right] \gamma ^{\alpha }\frac{i(p\hspace{-6pt}\slash-q^{\prime }\hspace{-7pt%
}\slash)+M}{M^{2}+(p-q^{\prime })^{2}}\gamma _{\mu }u_{\sigma }(p),
\end{eqnarray}%
where
\begin{eqnarray}
\mathcal{C}_{1}(q,M) &\simeq &\frac{2\left( \tau -1\right) }{M^{2}}\left(
\frac{M^{2}}{q^{2}}\right) ^{\tau -1}\left. _{2}F_{1}\right. (\tau -\frac{3}{%
2},\tau ,2\tau -3,-\frac{4M^{2}}{q^{2}}) \\
\mathcal{C}_{2}(q,M) &\simeq &\frac{2\tau }{M^{2}}\left( \frac{M^{2}}{q^{2}}%
\right) ^{\tau }\left. _{2}F_{1}\right. (\tau -\frac{1}{2},\tau +1,2\tau -1,-%
\frac{4M^{2}}{q^{2}}).
\end{eqnarray}%
To calculate the DVCS cross section, we define $T_{\mu \nu }=(2\pi
)^{4}\delta ^{(4)}(p+q-p^{\prime }-q^{\prime })\mathcal{M}_{\mu \nu }$. The
amplitude $\mathcal{M}_{\mu \nu }$ can be written as
\begin{eqnarray}
\mathcal{M}_{\mu \nu } &=&-\frac{\mathcal{Q}^{2}c_{i}^{\prime }c_{f}^{\prime
}}{z_{0}^{2}}e_{\nu \beta }^{L}(q)\bar{u}_{\sigma ^{\prime }}(p^{\prime
})\gamma _{\mu }\frac{i(p\hspace{-6pt}\slash+q\hspace{-6pt}\slash)+M}{%
M^{2}+(p+q)^{2}}\gamma ^{\beta }\left[ \mathcal{C}_{1}(q,M)P_{+}+\mathcal{C}%
_{2}(q,M)P_{-}\right] u_{\sigma }(p)  \nonumber \\
&&-\frac{\mathcal{Q}^{2}c_{i}^{\prime }c_{f}^{\prime }}{z_{0}^{2}}e_{\nu
\alpha }^{L}(q)\bar{u}_{\sigma ^{\prime }}(p^{\prime })\left[ \mathcal{C}%
_{1}(q,M)P_{-}+\mathcal{C}_{2}(q,M)P_{+}\right] \gamma ^{\alpha }\frac{i(p%
\hspace{-6pt}\slash-q^{\prime }\hspace{-7pt}\slash)+M}{M^{2}+(p-q^{\prime
})^{2}}\gamma _{\mu }u_{\sigma }(p),
\end{eqnarray}%
where one can see $\mathcal{M}_{\mu \nu }$ is dimensionless with our
normalization. One can further simplify the amplitude as
\begin{eqnarray}
\mathcal{M}_{\mu \nu } &=&\mathcal{D}_{1}\left[ \bar{u}_{\sigma ^{\prime
}}(p^{\prime })\gamma _{\mu }\frac{i(p\hspace{-6pt}\slash+q\hspace{-6pt}%
\slash)+M}{M^{2}+(p+q)^{2}}\gamma _{\nu }u_{\sigma }(p)+\bar{u}_{\sigma
^{\prime }}(p^{\prime })\gamma _{\nu }\frac{i(p\hspace{-6pt}\slash-q^{\prime
}\hspace{-7.5pt}\slash)+M}{M^{2}+(p+q^{\prime })^{2}}\gamma _{\mu }u_{\sigma
}(p)\right]  \nonumber \\
&&+\mathcal{D}_{2}e_{\nu \alpha }^{L}(q)\bar{u}_{\sigma ^{\prime
}}(p^{\prime })\gamma _{\mu }\frac{i(p\hspace{-6pt}\slash+q\hspace{-6.5pt}%
\slash)+M}{M^{2}+(p+q)^{2}}\gamma ^{\alpha }\gamma ^{5}u_{\sigma }(p)
\nonumber \\
&&-\mathcal{D}_{2}e_{\nu \alpha }^{L}(q)\bar{u}_{\sigma ^{\prime
}}(p^{\prime })\gamma ^{5}\gamma ^{\alpha }\frac{i(p\hspace{-6pt}\slash%
-q^{\prime }\hspace{-7.5pt}\slash)+M}{M^{2}+(p-q^{\prime })^{2}}\gamma _{\mu
}u_{\sigma }(p),
\end{eqnarray}%
with $\mathcal{D}_{1}=-\frac{\mathcal{Q}^{2}c_{i}^{\prime }c_{f}^{\prime }}{%
z_{0}^{2}}\frac{\mathcal{C}_{1}(q,M)+\mathcal{C}_{2}(q,M)}{2}$ and $\mathcal{%
D}_{2}=-\frac{\mathcal{Q}^{2}c_{i}^{\prime }c_{f}^{\prime }}{z_{0}^{2}}\frac{%
\mathcal{C}_{1}(q,M)-\mathcal{C}_{2}(q,M)}{2}$.

\subsubsection{Polarizations of photons}

Before we evaluate the DVCS cross section, we need to write down the
polarization of real and virtual photons in the lab frame. First for a real
photon, we suppose the momentum and the polarization of the photon can be
written as
\begin{eqnarray}
q_{\mu }^{\prime } &=&\left( \omega ^{\prime },\omega ^{\prime }\sin \theta
,0,\omega ^{\prime }\cos \theta \right) , \\
\varepsilon ^{\prime \mu }\left( q^{\prime },+1\right) &=&\left( 0,\cos
\theta ,0,-\sin \theta \right) , \\
\varepsilon ^{\prime \mu }\left( q^{\prime },-1\right) &=&\left(
0,0,1,0\right) .
\end{eqnarray}%
According to the current conservation%
\begin{equation}
\eta ^{\mu \nu }q_{\mu }^{\prime }\mathcal{M}_{\nu }=0,
\end{equation}%
one obtains
\begin{equation}
-\omega ^{\prime }\mathcal{M}_{0}+\omega ^{\prime }\mathcal{M}_{1}\sin
\theta +\omega ^{\prime }\mathcal{M}_{3}\cos \theta =0.
\end{equation}%
It is straightforward to show that
\begin{equation}
\eta ^{\mu \nu }\mathcal{M}_{\mu }^{\ast }\mathcal{M}_{\nu }=\left\vert
\mathcal{M}_{1}\right\vert ^{2}+\left\vert \mathcal{M}_{2}\right\vert
^{2}+\left\vert \mathcal{M}_{3}\right\vert ^{2}-\left\vert \mathcal{M}%
_{0}\right\vert ^{2}=\sum_{\lambda }\epsilon ^{\ast \mu }\epsilon ^{\nu }%
\mathcal{M}_{\mu }^{\ast }\mathcal{M}_{\nu }.
\end{equation}%
Therefore, we see that $\sum_{\lambda }\epsilon ^{\ast \mu }\epsilon ^{\nu
}=\eta ^{\mu \nu }$.

Now for a virtual photon, we find that the momentum and the polarization are
\begin{eqnarray}
q_{\mu } &=&\left( q_{0},0,0,q_{3}\right) , \\
\varepsilon ^{\mu }\left( q,0\right) &=&\frac{1}{q}\left(
q_{3},0,0,q_{0}\right) , \\
\varepsilon ^{\mu }\left( q,+1\right) &=&\left( 0,1,0,0\right) , \\
\varepsilon ^{\mu }\left( q,-1\right) &=&\left( 0,0,1,0\right) ,
\end{eqnarray}%
where $q^{2}=q_{3}^{2}-q_{0}^{2}$. According to the current conservation%
\begin{equation}
\eta ^{\mu \nu }q_{\mu }\mathcal{M}_{\nu }=0,
\end{equation}%
one gets
\begin{equation}
-q_{0}\mathcal{M}_{0}+q_{3}\mathcal{M}_{3}=0.
\end{equation}%
It is easy to see that
\begin{eqnarray}
\eta ^{\mu \nu }\mathcal{M}_{\mu }^{\ast }\mathcal{M}_{\nu } &=&\left\vert
\mathcal{M}_{1}\right\vert ^{2}+\left\vert \mathcal{M}_{2}\right\vert
^{2}+\left\vert \mathcal{M}_{3}\right\vert ^{2}-\left\vert \mathcal{M}%
_{0}\right\vert ^{2}  \nonumber \\
&=&\left\vert \mathcal{M}_{1}\right\vert ^{2}+\left\vert \mathcal{M}%
_{2}\right\vert ^{2}-\frac{q^{2}}{q_{0}^{2}}\left\vert \mathcal{M}%
_{3}\right\vert ^{2},
\end{eqnarray}%
and
\begin{eqnarray}
\sum_{\lambda }\epsilon ^{\ast \mu }\epsilon ^{\nu }\mathcal{M}_{\mu }^{\ast
}\mathcal{M}_{\nu } &=&\left\vert \mathcal{M}_{1}\right\vert ^{2}+\left\vert
\mathcal{M}_{2}\right\vert ^{2}+\left( -\frac{q_{3}}{q}\mathcal{M}_{0}^{\ast
}+\frac{q_{0}}{q}\mathcal{M}_{3}^{\ast }\right) \left( -\frac{q_{3}}{q}%
\mathcal{M}_{0}+\frac{q_{0}}{q}\mathcal{M}_{3}\right) \\
&=&\left\vert \mathcal{M}_{1}\right\vert ^{2}+\left\vert \mathcal{M}%
_{2}\right\vert ^{2}+\frac{q^{2}}{q_{0}^{2}}\left\vert \mathcal{M}%
_{3}\right\vert ^{2}.
\end{eqnarray}%
Therefore, for a virtual photon, we know
\begin{equation}
\sum_{\lambda }\epsilon ^{\ast \mu }\epsilon ^{\nu }\neq \eta ^{\mu \nu },%
\text{ but }\sum_{\lambda }\left( -1\right) ^{\lambda +1}\epsilon ^{\ast \mu
}\epsilon ^{\nu }=\eta ^{\mu \nu }.
\end{equation}

\subsubsection{DVCS cross sections}

Using the results from the last subsection, we can find the following
identity
\begin{eqnarray}
2\left(\sigma_{T}+\sigma_{L}\right) &=&\left\vert
\mathcal{M}_{1}\right\vert
^{2}+\left\vert \mathcal{M}_{2}\right\vert ^{2}+2\frac{q^{2}}{q_{0}^{2}}\left\vert \mathcal{M}_{3}\right\vert ^{2}, \\
&=&\eta ^{\mu \nu }\mathcal{M}_{\mu }^{\ast }\mathcal{M}_{\nu }+3\frac{q^{2}%
}{q_{3}^{2}M^{2}}p^{\mu }p^{\nu }\mathcal{M}_{\mu }^{\ast
}\mathcal{M}_{\nu},.
\end{eqnarray}%
Together with the definition
\begin{equation}
W_{\nu \nu ^{\prime }}=\varepsilon ^{\mu }(q^{\prime })\varepsilon ^{\ast
\mu ^{\prime }}(q^{\prime })\mathcal{M}_{\mu \nu }\mathcal{M}_{\mu ^{\prime
}\nu ^{\prime }}^{\ast },  \label{wab}
\end{equation}%
after some tedious but straightforward computation, we find
\[
\sum_{\lambda }\epsilon ^{\ast \mu }\epsilon ^{\nu }W_{\mu \nu }=8\mathcal{D}%
_{1}\mathcal{D}_{1}\left\{ A_{ss}^{\left( 1\right) }+A_{uu}^{\left( 1\right)
}+A_{us}^{\left( 1\right) }\right\} +8\mathcal{D}_{2}\mathcal{D}_{2}\left\{
A_{ss}^{\left( 2\right) }+A_{uu}^{\left( 2\right) }+A_{us}^{\left( 2\right)
}\right\} ,
\]%
with
\begin{eqnarray}
A_{ss}^{\left( 1\right) } &=&\frac{1}{\tilde{s}^{2}}(4M^{4}-4M^{2}\tilde{s}%
-2M^{2}t-2M^{2}\tilde{u}-\tilde{s}\tilde{u})  \nonumber \\
&&+\frac{3q^{2}}{q_{3}^{2}\tilde{s}^{2}}\left( M^{2}t+4M^{4}-M^{2}\tilde{s}%
+M^{2}\tilde{u}+\frac{1}{2}\tilde{u}\tilde{s}\right) \\
A_{uu}^{\left( 1\right) } &=&\frac{1}{\tilde{u}^{2}}(4M^{4}-2M^{2}\tilde{s}%
-2M^{2}t-4M^{2}\tilde{u}-\tilde{s}\tilde{u})  \nonumber \\
&&+\frac{3q^{2}}{q_{3}^{2}\tilde{u}^{2}}\left[ (t+2M^{2})(M^{2}-\tilde{u}-%
\frac{\tilde{u}^{2}}{2M^{2}})+2M^{4}-\frac{1}{2}\tilde{u}\tilde{s}-M^{2}(%
\tilde{s}+\tilde{u})\right] \\
A_{us}^{\left( 1\right) } &=&\frac{2}{\tilde{u}\tilde{s}}\left[ M^{2}(4M^{2}-%
\tilde{s}-\tilde{u})-t(t+\tilde{s}+\tilde{u})\right]  \nonumber \\
&&+\frac{3q^{2}}{q_{3}^{2}\tilde{u}\tilde{s}}\left[ (t+2M^{2})(4M^{2}+\tilde{%
u}+t)-\tilde{u}^{2}+2M^{2}\tilde{s}\right] ,
\end{eqnarray}%
and
\begin{eqnarray}
&&A_{ss}^{\left( 1\right) }+A_{uu}^{\left( 1\right) }+A_{us}^{\left(
1\right) }  \nonumber \\
&=&-\frac{1}{\tilde{s}^{2}\tilde{u}^{2}}\left\{ 2\tilde{s}\tilde{u}\left[
t(t+\tilde{s}+\tilde{u})-M^{2}(4M^{2}-\tilde{s}-\tilde{u})\right] \right.
\nonumber \\
&&-\tilde{u}^{2}(4M^{4}-4M^{2}\tilde{s}-2M^{2}t-2M^{2}\tilde{u}-\tilde{s}%
\tilde{u})  \nonumber \\
&&\left. -\tilde{s}^{2}(4M^{4}-4M^{2}\tilde{u}-2M^{2}t-2M^{2}\tilde{s}-%
\tilde{s}\tilde{u})\right\}  \nonumber \\
&&+\frac{3q^{2}}{q_{3}^{2}\tilde{s}^{2}\tilde{u}^{2}}\left( 4M^{2}-\tilde{s}%
+t+\tilde{u}-\frac{\tilde{s}\tilde{u}}{2M^{2}}\right) \left[ M^{2}(\tilde{s}+%
\tilde{u})^{2}+t\tilde{s}\tilde{u}\right] \\
&=&-\frac{1}{\tilde{s}^{2}\tilde{u}^{2}}\left\{ \left( \tilde{s}^{2}+\tilde{u%
}^{2}\right) \tilde{s}\tilde{u}+2\left( q^{2}-2M^{2}\right) \left[
M^{2}\left( \tilde{s}+\tilde{u}\right) ^{2}+\tilde{u}\tilde{s}t\right]
\right\}  \nonumber \\
&&+\frac{3q^{2}}{q_{3}^{2}\tilde{s}^{2}\tilde{u}^{2}}\left( 4M^{2}-\tilde{s}%
+t+\tilde{u}-\frac{\tilde{s}\tilde{u}}{2M^{2}}\right) \left[ M^{2}(\tilde{s}+%
\tilde{u})^{2}+t\tilde{s}\tilde{u}\right] ,
\end{eqnarray}%
and
\begin{eqnarray}
A_{ss}^{\left( 2\right) } &=&-\frac{1}{2q^{2}\tilde{s}^{2}}\left[
2M^{2}(8M^{2}q^{2}+2q^{4}-4q^{2}\tilde{s}-\tilde{s}^{2})-\tilde{s}^{2}(%
\tilde{s}+\tilde{u})+q^{2}\tilde{s}(\tilde{s}+2\tilde{u})\right]  \nonumber
\\
&&-\frac{3}{8q^{2}q_{3}^{2}}\left[
4M^{2}q^{2}+(q^{2}-\tilde{s})^{2}\right]
(-2+\frac{t}{M^{2}}) \\
A_{uu}^{\left( 2\right) } &=&-\frac{1}{2q^{2}\tilde{u}^{2}}\left[
2M^{2}(8M^{2}q^{2}+2q^{4}-4q^{2}\tilde{u}-\tilde{u}^{2})-\tilde{u}^{2}(%
\tilde{s}+\tilde{u})+q^{2}\tilde{u}(\tilde{u}+2\tilde{s})\right]  \nonumber
\\
&&-\frac{3}{8q^{2}q_{3}^{2}\tilde{u}^{2}}\left[ 2(4M^{2}q^{2}\tilde{s}%
^{2}+8M^{2}q^{2}\tilde{s}\tilde{u}+4q^{4}\tilde{s}\tilde{u}+q^{4}\tilde{u}%
^{2}-2q^{2}\tilde{s}^{2}\tilde{u}-\tilde{s}^{2}\tilde{u}^{2})\right.
\nonumber \\
&&\left. \hspace{2cm}+\frac{t}{M^{2}}(q^{2}+\tilde{s})^{2}\tilde{u}^{2}%
\right] \\
A_{us}^{\left( 2\right) } &=&-\frac{1}{q^{2}\tilde{u}\tilde{s}}\left[
4M^{2}q^{2}(4M^{2}+2q^{2}+t)-2M^{2}(\tilde{s}^{2}+\tilde{u}^{2}+\tilde{s}%
\tilde{u})+(2q^{4}-\tilde{s}\tilde{u})t\right]  \nonumber \\
&&-\frac{3}{4q^{2}q_{3}^{2}\tilde{u}}\left[ 2(4M^{2}q^{2}\tilde{u}+3q^{4}%
\tilde{u}+q^{2}\tilde{s}^{2}-2q^{2}\tilde{s}\tilde{u}-q^{2}\tilde{u}^{2}-%
\tilde{s}^{3}-\tilde{s}^{2}\tilde{u}-\tilde{s}\tilde{u}^{2})\right.
\nonumber \\
&&\left. \hspace{2cm}+\frac{t}{M^{2}}(q^{4}-\tilde{s}^{2})\tilde{u}\right]
\end{eqnarray}%
and
\begin{eqnarray}
&&A_{ss}^{\left( 2\right) }+A_{uu}^{\left( 2\right) }+A_{us}^{\left(
2\right) }  \nonumber \\
&=&-\frac{1}{q^{2}\tilde{s}^{2}\tilde{u}^{2}}\left[ 2M^{2}(\tilde{s}+\tilde{u%
})^{2}(4M^{2}q^{2}+q^{4}-\tilde{s}\tilde{u})+q^{2}\tilde{s}\tilde{u}(\tilde{s%
}^{2}+\tilde{u}^{2}+8M^{2}t+2q^{2}t)\right]  \nonumber \\
&&-\frac{3}{2q^{2}q_{3}^{2}\tilde{u}^{2}}\left[ (2M^{2}q^{2}-\tilde{s}\tilde{u%
})(\tilde{s}+\tilde{u})^{2}+2q^{4}\tilde{u}(\tilde{s}+\tilde{u})+q^{2}\tilde{%
u}^{2}t(2+q^{2}/M^{2})\right] .
\end{eqnarray}%
The Mandelstam variables used above are defined as follows
\begin{eqnarray}
\tilde{s} &=&s+M^{2}=(p^{\prime }+q^{\prime })^{2}+M^{2}=2p^{\prime }\cdot
q^{\prime }, \\
\tilde{u} &=&u+M^{2}=(p-q^{\prime })^{2}+M^{2}=-2p\cdot q^{\prime }, \\
t &=&(p-p^{\prime })^{2}.
\end{eqnarray}

Using the lab frame we define in Eqs.(\ref{frame1}) and Eq.(\ref{frame2}),
along with the phase space integral
\begin{equation}
\int d\Pi _{2}==\frac{1}{8\pi \sqrt{4M^{2}q^{2}+(W^{2}+q^{2}-M^{2})^{2}}}%
\int dt,\quad \text{with}\quad W^{2}=2Mq_{0}-q^{2}+M^{2},
\end{equation}%
we can write cross-section formula as
\begin{eqnarray}
\frac{d\sigma ^{\text{Un}}}{dt}
&=&\frac{1}{2q_{0}}\frac{1}{2M}\frac{1}{4\pi
}\frac{\omega ^{\prime 2}}{2Mq_{0}-q^{2}}\frac{1}{4}2\left(\sigma_{T}+\sigma_{L}\right)  \nonumber \\
&=&\frac{1}{16\pi (W^{2}+q^{2}-M^{2})\sqrt{%
4M^{2}q^{2}+(W^{2}+q^{2}-M^{2})^{2}}}  \nonumber \\
&&\times \left[ 2\mathcal{D}_{1}^{2}\left( A_{ss}^{\left( 1\right)
}+A_{uu}^{\left( 1\right) }+A_{us}^{\left( 1\right) }\right) +2\mathcal{D}%
_{2}^{2}\left( A_{ss}^{\left( 2\right) }+A_{uu}^{\left( 2\right)
}+A_{us}^{\left( 2\right) }\right) \right] .  \label{fdvcs}
\end{eqnarray}

\begin{figure}[tbp]
\begin{center}
\includegraphics[width=15cm]{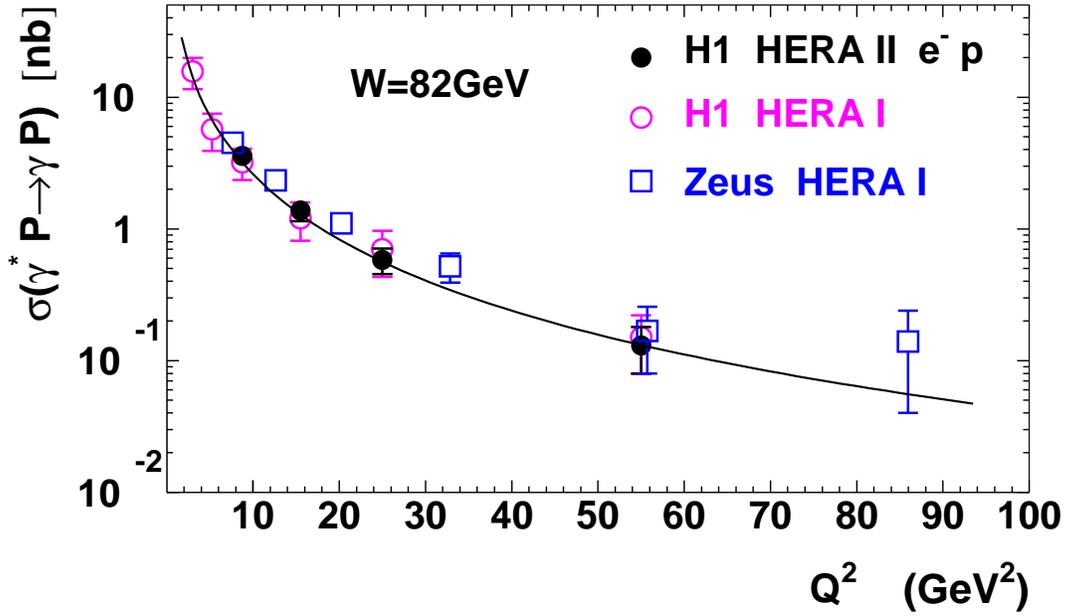}
\end{center}
\caption[*]{The DVCS ($\protect\gamma^{*}p\rightarrow \protect\gamma p$)
cross section as a function of $Q^2$ for $W=82 GeV$ and $|t|<1 GeV^2$. The
data points are taken from Refs.~\protect\cite{Aktas:2005ty, Chekanov:2003ya, :2007cz}%
. The solid line represents the dilatino DVCS cross section in Eq.~(\protect\ref{fdvcs})
after integrating over $t$. We set $\tau=2$, the mass $M=0.938 GeV$ to be the proton mass and $q^2=Q^2$. We have
used one of the data points to fix the overall constant.  }
\label{Sigma}
\end{figure}
The above equation is one of our major results in this paper.
Assuming the QCD coupling is large during the interaction due to confinement,
we can compare this result to the HERA data.
We plot the integrated cross section as a function of $Q^2$ in
Fig.~\ref{Sigma}. As shown in Fig.~\ref{Sigma}, we compare the
DVCS cross section in Eq.~(\ref{fdvcs}) after integrating over $t$
($|t|<1$) to the low-energy H1 and the ZEUS
data\cite{Aktas:2005ty, Chekanov:2003ya, :2007cz} and find that our result
is consistent with the data. Equation~(\ref{fdvcs}) only includes the
\textit{s}-channel and \textit{u}-channel graphs. There should be the \textit{t}-channel
graviton exchange contribution which is discussed in the next
section in the plot. However, we argue that the \textit{s}-channel and
\textit{u}-channel contributions are dominant in the low-energy region,
whereas the \textit{t}-channel graviton exchange dominates in the high-energy limit.
 In addition, technically, it is hard to fix the
normalization of the \textit{t}-channel graph as compared to the \textit{s}-channel and
\textit{u}-channel graphs. It is obvious
that the \textit{s}-channel and \textit{u}-channel contributions are proportional to $\mathcal{%
Q}^2$, with $\mathcal{Q}$ being the charge of the targets. On the other
hand, the \textit{t}-channel graviton exchange only couples to the energy momentum
tensors of the target hadrons, and it does not depend on the quantum number
of the target. In QCD, the situation is quite similar. In the low-energy
region, where valence quark contribution dominates, the scattering amplitude
certainly is proportional to $e^2$, where $e$ is the proton charge. However,
in the high-energy region, where the Pomeron exchange starts to dominate the
amplitude, we find the parton distribution becomes universal and independent
of the quantum number of the target hadron. Therefore, assuming the
\textit{t}-channel graviton exchange graphs are small in the low-energy region, we
neglect this contribution when we compare the calculated cross section to
the low-energy H1 and the ZEUS data\cite{Aktas:2005ty, Chekanov:2003ya, :2007cz}.

\begin{figure}[tbp]
\begin{center}
\includegraphics[width=15cm]{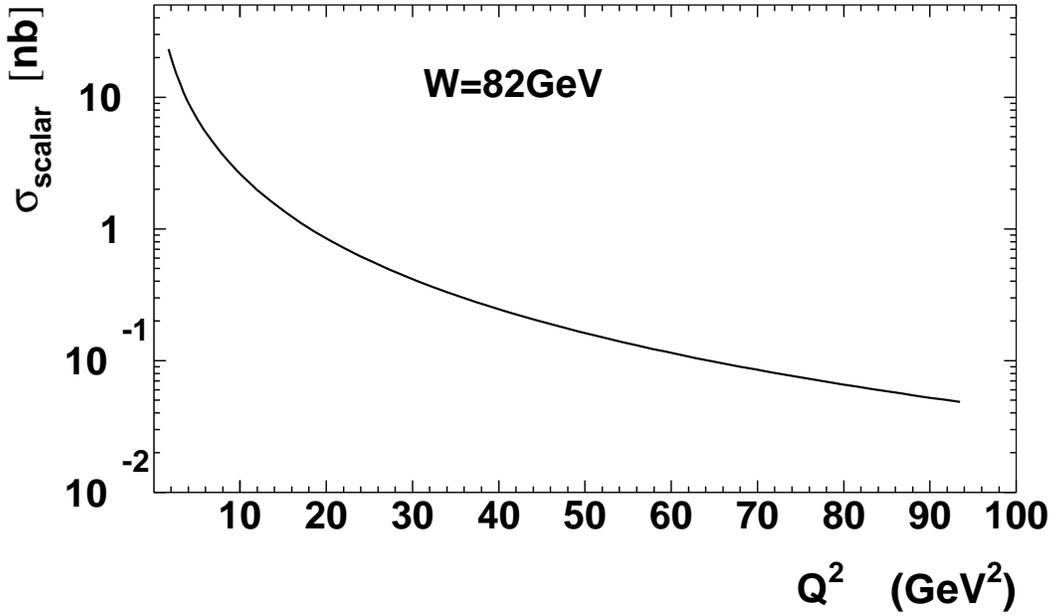}
\end{center}
\caption[*]{The scalar DVCS cross section as a function of $Q^2$ for $W=82
GeV$ and $|t|<1 GeV^2$. The solid line represents the dilaton DVCS cross section
calculated in Sec.~\ref{scalar}. We set $\Delta =2$ and use the same normalization and parameters as the dilatino in the plot.}
\label{Sigma2}
\end{figure}

As a comparison, we also plot the integrated cross section for a scalar
target, which is calculated in Sec.~\ref{scalar}, as function of $Q^2$ in
Fig.~\ref{Sigma2}. We use the same parameter and overall constant as in
Fig.~\ref{Sigma}. Numerically, there is little difference between these
two plots.

\section{\textit{T}-channel graviton exchanges}

\label{gravitonex}
\begin{figure}[tbp]
\begin{center}
\includegraphics[width=6cm]{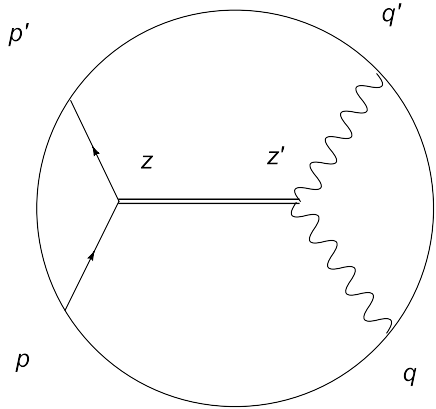}
\end{center}
\caption[*]{\textit{t}-channel graviton exchange}
\label{DVCS4}
\end{figure}

The exact evaluation of the \textit{t}-channel graviton exchange graph as illustrated
in Fig.~\ref{DVCS4} is not an easy task. Here we provide a heuristic
derivation at the large-$s$ limit where the \textit{t}-channel graviton exchange
contribution dominates, with $s$ being the center of mass energy. Therefore,
hereafter we neglect the mass terms in the energy momentum tensors and only
keep the leading order terms. For simplicity, we also drop the $S^{5}$
dependence of the wave functions, because it only contributes an overall
constant. It is well-known that the graviton couples to the energy momentum tensor.
The \textit{t}-channel graviton exchange contribution does not depend on the charge
of the target hadron ($\mathcal{Q}$), which is uniquely different from the
\textit{s}-channel and \textit{u}-channel graphs' contributions computed above. Technically, it
is hard to normalize the \textit{t}-channel graviton contribution as compared to the
\textit{s}-channel and \textit{u}-channel graphs' contributions. We assume that the
\textit{t}-channel contribution is small at low energy. However, it is the dominant
contribution at high energy due to its energy dependence. In this section,
we set the curvature of the AdS space $R=1$, since the final results do
not depend on $R$.

The \textit{t}-channel graviton exchange amplitude between the dilaton target and the
photon can be written in the following:
\begin{equation}
\mathcal{A}^{G}=n^{\mu }T_{\mu \nu }^{G}n^{\prime \ast \nu }=\kappa ^{2}\int
d^{5}yd^{5}y^{\prime }T_{mn}^{\Phi }(y)G^{mnkl}(y,y^{\prime
})T_{kl}^{A}(y^{\prime }),
\end{equation}%
where $n^{\mu }$ and $n^{\prime \ast \nu }$ are the polarization vectors and
$\kappa $ is the gravitational coupling to the stress-energy tensor. $%
G^{mnkl}(x,y)$ is the graviton propagator which can be written as\cite%
{Polchinski:1998rq}
\begin{equation}
G^{mnkl}(y,y^{\prime })=\left( g^{mk}g^{nl}+g^{ml}g^{nk}-\eta
g^{mn}g^{kl}\right) G(y,y^{\prime })
\end{equation}%
with $\eta =\frac{2}{D-2}$. Strictly speaking, one should use the full
graviton propagator derived in $AdS_{5}$ space as in Ref. \cite%
{D'Hoker:1999pj}. However, in the high-energy limit, one can still obtain
the correct expression using the above graviton propagator (see, e.g., in Refs.
\cite{Polchinski:2002jw, Bartels:2009sc}). The stress-energy tensor $%
T_{mn}^{\Phi }$ for a massless dilaton field is
\begin{equation}
T_{mn}^{\Phi }=\left( g_{m\alpha }g_{n\beta }+g_{m\beta }g_{n\alpha
}-g_{mn}g_{\alpha \beta }\right) \partial ^{\alpha }\Phi ^{\ast }\partial
^{\beta }\Phi .
\end{equation}%
We put $\Phi $ as the initial dilaton wave function and $\Phi ^{\ast }$ as
the final dilaton wave function. The stress-energy tensor $T_{kl}^{A}$ for a
Kaluza-Klein gauge field can be written as
\begin{equation}
T_{kl}^{A}=g^{\rho \sigma }F_{k\rho }F_{l\sigma }-\frac{1}{4}g_{kl}F_{\rho
\sigma }F^{\rho \sigma },
\end{equation}%
with
\begin{eqnarray}
F_{\mu \nu } &=&i\left( q_{\mu }n_{\nu }-q_{\nu }n_{\mu }\right) e^{iq\cdot
y}qzK_{1}\left( qz\right) ,  \nonumber \\
F_{\mu z} &=&\left( n_{\mu }q^{2}-q_{\mu }q\cdot n\right) e^{iq\cdot
y}zK_{0}\left( qz\right) .
\end{eqnarray}%
The indices $\mu ,\nu $ denote the four dimensions in Minkowski spacetime,
and the rest of the indices $m,n,k,l,\cdots $ are defined in $\textrm{AdS}_{5}$
space. It is easy to work out the tensor contractions and obtain
\begin{eqnarray}
&&T_{mn}^{\Phi }(y)G^{mnkl}(y,y^{\prime })T_{kl}^{A}(y^{\prime })  \nonumber
\\
&\propto &2\left( \partial ^{k}\Phi ^{\ast }\partial ^{l}\Phi +\partial
^{k}\Phi \partial ^{l}\Phi ^{\ast }\right) F_{k\rho }F_{l\sigma }g^{\rho
\sigma }-\partial ^{k}\Phi \partial _{k}\Phi F_{\rho \sigma }F^{\rho \sigma
}+\cdots .
\end{eqnarray}%
This expression agrees with Ref. \cite{Polchinski:2002jw} in the forward
limit. At high energy\cite{Bartels:2009sc}, the first term in the above
effective action dominates, because it gives rise to terms like $p\cdot q$,
which is large as compared to $q^{2}$ and $M^{2}$. Therefore, the graviton
exchange amplitude for the Compton scattering can be written as
\begin{equation}
\mathcal{A}^{G}=n^{\mu }{\mathcal{T}}_{\mu \nu }^{G}n^{\prime \ast \nu
}=\int d^{5}y\sqrt{-g}d^{5}y^{\prime }\sqrt{-g^{\prime }}2\left( \partial
^{k}\Phi ^{\ast }\partial ^{l}\Phi +\partial ^{k}\Phi \partial ^{k}\Phi
^{\ast }\right) g^{\rho \sigma }F_{k\rho }F_{l\sigma }G(y,y^{\prime }),
\end{equation}%
It is straightforward to find that
\begin{eqnarray}
&&g^{\rho \sigma }F_{\mu \rho }\left( q\right) F_{\nu \sigma }\left(
q^{\prime }\right)   \nonumber \\
&=&\left( n_{\mu }q^{2}-q_{\mu }q\cdot n\right) \left( n_{\nu }^{\prime
}q^{\prime 2}-q_{\nu }^{\prime }q^{\prime }\cdot n^{\prime }\right) \frac{%
z^{4}}{R^{2}}e^{i\left( q-q^{\prime }\right) \cdot y}K_{0}\left( qz\right)
K_{0}\left( q^{\prime }z\right)   \nonumber \\
&&+\left( n\cdot n^{\prime }q_{\mu }q_{\nu }^{\prime }-q\cdot n^{\prime
}n_{\mu }q_{\nu }^{\prime }-q^{\prime }\cdot nq_{\mu }n_{\nu }^{\prime
}+q\cdot q^{\prime }n_{\mu }n_{\nu }^{\prime }\right) \frac{z^{4}}{R^{2}}%
e^{i\left( q-q^{\prime }\right) \cdot y}q^{\prime }qK_{1}\left( qz\right)
K_{1}\left( q^{\prime }z\right) .
\end{eqnarray}%
Following the derivation in Ref.\cite{Hatta:2007he}, we get
\begin{eqnarray}
{\mathcal{T}}_{\mu \nu }^{G} &\propto &\mathcal{A}^{\prime \prime }\left[
\left( p_{\mu }+\frac{q_{\mu }}{2x}\right) \left( p_{\nu }^{\prime }+\frac{%
q_{\nu }^{\prime }}{2x^{\prime }}\right) +\left( p_{\mu }^{\prime }-\frac{%
q_{\mu }}{2\tilde{x}}\right) \left( p_{\nu }-\frac{q_{\nu }^{\prime }}{2%
\tilde{x}^{\prime }}\right) \right] q^{2}q^{\prime 2}\mathcal{T}_{1}^{G}
\nonumber \\
&&+\mathcal{A}^{\prime \prime }\left[
\begin{array}{c}
\frac{q^{\prime }q}{4xx^{\prime }}\left( \eta _{\mu \nu }+\frac{2x}{q^{2}}%
p_{\mu }q_{\nu }+\frac{2x^{\prime }}{q^{\prime 2}}q_{\mu }^{\prime }p_{\nu
}^{\prime }+\frac{4xx^{\prime }}{q^{2}q^{\prime 2}}q\cdot q^{\prime }p_{\mu
}p_{\nu }^{\prime }\right)  \\
+\frac{q^{\prime }q}{4\tilde{x}\tilde{x}^{\prime }}\left( \eta _{\mu \nu }-%
\frac{2\tilde{x}}{q^{2}}p_{\mu }^{\prime }q_{\nu }-\frac{2\tilde{x}^{\prime }%
}{q^{\prime 2}}q_{\mu }^{\prime }p_{\nu }+\frac{4\tilde{x}\tilde{x}^{\prime }%
}{q^{2}q^{\prime 2}}q\cdot q^{\prime }p_{\mu }^{\prime }p_{\nu }\right)
\end{array}%
\right] q^{2}q^{\prime 2}\mathcal{T}_{2}^{G},
\end{eqnarray}%
with $\mathcal{A}^{\prime \prime }\propto \left( 2\pi \right) ^{4}\delta
\left( p+q-p^{\prime }-q^{\prime }\right) $, $x=-\frac{q^{2}}{2p\cdot q}$, $%
x^{\prime }=-\frac{q^{\prime 2}}{2p^{\prime }\cdot q^{\prime }}$, $\tilde{x}=%
\frac{q^{2}}{2p^{\prime }\cdot q}$, $\tilde{x}^{\prime }=\frac{q^{\prime 2}}{%
2p\cdot q^{\prime }}$ and
\begin{eqnarray}
\mathcal{T}_{1}^{G} &=&\int dzz^{3}K_{0}\left( qz\right) K_{0}\left(
q^{\prime }z\right) \int dz^{\prime }z^{\prime 3}J_{\Delta -2}^{2}\left(
Mz^{\prime }\right)   \nonumber \\
&&\times \int_{\mathcal{C}}\frac{dj}{2i}\frac{1+e^{-i\pi j}}{\sin \pi j}%
\frac{\left( \alpha ^{\prime }s\right) ^{j-2+\alpha ^{\prime }tz^{\prime 2}}%
}{\sqrt{j-j_{0}}}\left( \frac{z^{2}}{z^{\prime 2}}\right) ^{1-\gamma }, \\
\mathcal{T}_{2}^{G} &=&\int dzz^{3}K_{1}\left( qz\right) K_{1}\left(
q^{\prime }z\right) \int dz^{\prime }z^{\prime 3}J_{\Delta -2}^{2}\left(
Mz^{\prime }\right)   \nonumber \\
&&\times \int_{\mathcal{C}}\frac{dj}{2i}\frac{1+e^{-i\pi j}}{\sin \pi j}%
\frac{\left( \alpha ^{\prime }s\right) ^{j-2+\alpha ^{\prime }tz^{\prime 2}}%
}{\sqrt{j-j_{0}}}\left( \frac{z^{2}}{z^{\prime 2}}\right) ^{1-\gamma },
\end{eqnarray}%
with $\alpha ^{\prime }=R^{2}/\sqrt{\lambda }$ and $\gamma \simeq 1-\frac{1}{%
\sqrt{\lambda }}$. Here we have neglected the contributions of $g^{\rho
\sigma }F_{z\rho }\left( q\right) F_{\nu \sigma }\left( q^{\prime }\right) $
and $g^{\rho \sigma }F_{z\rho }\left( q\right) F_{z\sigma }\left( q^{\prime
}\right) $ since they are subleading at the high-energy limit. The contour $%
\mathcal{C}$ runs parallel to the imaginary axis from $j-i\infty $ to $%
j+i\infty $ with $j_{0}\equiv 2-\frac{2}{\sqrt{\lambda }}<j<2$. It is
interesting to note that $\mathcal{T}_{\mu \nu }^{G}$ is always conserved,
namely, $q^{\mu }\mathcal{T}_{\mu \nu }^{G}=0$ and $q^{\prime \nu }\mathcal{T}%
_{\mu \nu }^{G}=0$, thanks to its coupling to the photon energy momentum
tensor and the equation of motion.

Assuming that the interaction is local\footnote{%
Essentially, the \textit{t}-channel graviton exchange is non-local. Here we adopt
this local approximation  to show that the imaginary part of this
amplitude in the forward limit can smoothly match the DIS calculation\cite%
{Polchinski:2002jw, Hatta:2007he}.}, one obtains%
\begin{eqnarray}
\mathcal{T}_{1}^{G} &=&\int dzz^{3}K_{0}\left( qz\right) K_{0}\left(
q^{\prime }z\right) z^{4}J_{\Delta -2}^{2}\left( Mz\right) ,  \nonumber \\
\mathcal{T}_{2}^{G} &=&\int dzz^{3}K_{1}\left( qz\right) K_{1}\left(
q^{\prime }z\right) z^{4}J_{\Delta -2}^{2}\left( Mz\right) .
\end{eqnarray}%
In the forward Compton scattering limit, our result agrees with the one
found in Refs.\cite{Polchinski:2002jw, Hatta:2007he}. If the energy gets so
high that $\frac{1}{\sqrt{\lambda }}\ln \alpha ^{\prime }s$ becomes of order
$1$, the imaginary part of the graviton exchange amplitude can no longer be
assumed local due to diffusion\cite{Polchinski:2002jw, Hatta:2007he}.
Because graviton has a spin of 2, the forward scattering amplitude is
proportional to $\frac{1}{x^{2}}$ up to some curvature corrections.

For a dilatino target, we use the following dilatino energy momentum tensor
\begin{equation}
T_{\Psi }^{mn}=\frac{1}{2}\left( \overline{\Psi }^{\ast }\gamma ^{m}%
\overrightarrow{\partial ^{n}}\Psi +\overline{\Psi }^{\ast }\gamma ^{m}%
\overleftarrow{\partial ^{n}}\Psi +\overline{\Psi }^{\ast }\gamma ^{n}%
\overrightarrow{\partial ^{m}}\Psi +\overline{\Psi }^{\ast }\gamma ^{n}%
\overleftarrow{\partial ^{m}}\Psi \right) .
\end{equation}%
From the above calculation for the dilaton, one can easily see that the leading
contribution at high energy comes from
\begin{equation}
T_{\Psi }^{\mu \nu }\propto 2\left( p^{\mu }p^{\prime \nu }+p^{\nu
}p^{\prime \mu }\right) e^{-i\left( p-p^{\prime }\right) \cdot x}z^{5}\left(
J_{\tau -2}^{2}\left( Mz^{\prime }\right) +J_{\tau -1}^{2}\left( Mz^{\prime
}\right) \right) +\cdots .
\end{equation}%
Thus the \textit{t}-channel graviton exchange amplitude for the dilatino is
\begin{eqnarray}
{\mathcal{T}}_{\mu \nu }^{G\Psi } &\propto &\mathcal{A}^{\prime \prime }%
\left[ \left( p_{\mu }+\frac{q_{\mu }}{2x}\right) \left( p_{\nu }^{\prime }+%
\frac{q_{\nu }^{\prime }}{2x^{\prime }}\right) +\left( p_{\mu }^{\prime }-%
\frac{q_{\mu }}{2\tilde{x}}\right) \left( p_{\nu }-\frac{q_{\nu }^{\prime }}{%
2\tilde{x}^{\prime }}\right) \right] q^{2}q^{\prime 2}\mathcal{T}_{1}^{G\Psi
}  \nonumber \\
&&+\mathcal{A}^{\prime \prime }\left[
\begin{array}{c}
\frac{q^{\prime }q}{4xx^{\prime }}\left( \eta _{\mu \nu }+\frac{2x}{q^{2}}%
p_{\mu }q_{\nu }+\frac{2x^{\prime }}{q^{\prime 2}}q_{\mu }^{\prime }p_{\nu
}^{\prime }+\frac{4xx^{\prime }}{q^{2}q^{\prime 2}}q\cdot q^{\prime }p_{\mu
}p_{\nu }^{\prime }\right) \\
+\frac{q^{\prime }q}{4\tilde{x}\tilde{x}^{\prime }}\left( \eta _{\mu \nu }-%
\frac{2\tilde{x}}{q^{2}}p_{\mu }^{\prime }q_{\nu }-\frac{2\tilde{x}^{\prime }%
}{q^{\prime 2}}q_{\mu }^{\prime }p_{\nu }+\frac{4\tilde{x}\tilde{x}^{\prime }%
}{q^{2}q^{\prime 2}}q\cdot q^{\prime }p_{\mu }^{\prime }p_{\nu }\right)%
\end{array}%
\right] q^{2}q^{\prime 2}\mathcal{T}_{2}^{G\Psi },
\end{eqnarray}%
with
\begin{eqnarray}
\mathcal{T}_{1}^{G\Psi } &=&\int dzz^{3}K_{0}\left( qz\right) K_{0}\left(
q^{\prime }z\right) \int dz^{\prime }z^{\prime 3}\left( J_{\tau
-2}^{2}\left( Mz^{\prime }\right) +J_{\tau -1}^{2}\left( Mz^{\prime }\right)
\right)  \nonumber \\
&&\times \int_{\mathcal{C}}\frac{dj}{2i}\frac{1+e^{-i\pi j}}{\sin \pi j}%
\frac{\left( \alpha ^{\prime }s\right) ^{j-2+\alpha ^{\prime }tz^{\prime 2}}%
}{\sqrt{j-j_{0}}}\left( \frac{z^{2}}{z^{\prime 2}}\right) ^{1-\gamma }, \\
\mathcal{T}_{2}^{G\Psi } &=&\int dzz^{3}K_{1}\left( qz\right) K_{1}\left(
q^{\prime }z\right) \int dz^{\prime }z^{\prime 3}\left( J_{\tau
-2}^{2}\left( Mz^{\prime }\right) +J_{\tau -1}^{2}\left( Mz^{\prime }\right)
\right)  \nonumber \\
&&\times \int_{\mathcal{C}}\frac{dj}{2i}\frac{1+e^{-i\pi j}}{\sin \pi j}%
\frac{\left( \alpha ^{\prime }s\right) ^{j-2+\alpha ^{\prime }tz^{\prime 2}}%
}{\sqrt{j-j_{0}}}\left( \frac{z^{2}}{z^{\prime 2}}\right) ^{1-\gamma }.
\end{eqnarray}%
The evaluation of the above Compton scattering amplitude is quite complicated.
Nevertheless, we can approximately find the corresponding DVCS cross section
for both dilaton and dilatino targets at the $t=0$ limit is
\begin{equation}
\frac{d\sigma _{\text{G}}^{\text{Un}}}{dt}\propto \frac{1}{s^{2}}\left( s^{2-%
\frac{2}{\sqrt{\lambda }}}\right) ^{2}\left( \frac{M^{2}}{Q^{2}}\right) ^{2},
\label{gravs}
\end{equation}%
where the factor $\frac{1}{s^{2}}$ comes from the phase space integral and $%
s^{2-\frac{2}{\sqrt{\lambda }}}$ arises due to the graviton exchange
together with its curvature correction. The energy enhancement in Eq.~(\ref%
{gravs}) indicates that the \textit{t}-channel graviton exchange should dominate the
total scattering amplitude in the high-energy limit.

\section{Conclusion}

\label{conclusion}

In this paper, we calculate the supergravity graphs corresponding to Compton
scatterings for both dilaton and dilatino targets in the CFT. For a dilaton
target, we have calculated the \textit{s}-channel, \textit{u}-channel graphs where all the
Kaluza-Klein excitations of intermediate states are summed over, and the
four-point contact term. We sum all these three graphs and compute the usual
real Compton scattering amplitude and DVCS cross section. For a dilatino
target, we compute the contributions from the \textit{s}-channel and \textit{u}-channel graphs
explicitly. We find that the real Compton scattering amplitude, up to an
overall constant, is identical to the amplitude found in scalar QED. The
DVCS cross section is also discussed afterwards and compared to the
low-energy H1 and ZEUS data. The curve is consistent with the experimental
data. The end of this paper is devoted to discussing the \textit{t}-channel graviton
exchange contribution, which is dominant contribution in the high-energy
limit. We have provided a heuristic derivation for the computation of the
\textit{t}-channel graviton exchange.

It will be interesting to continue this study and extract the generalized
parton distributions as well as the information about the total spin of
constituents in a hadron. This would help us to understand the role of the
orbital angular momentum and the spin sum rule at strongly coupled regime.

\begin{acknowledgments}
We acknowledge inspiring discussions with E. Iancu, Y. Hatta, C.
Marquet, A. H. Mueller and F. Yuan. J.~G.~
acknowledges financial support by the China Postdoctoral Science Foundation funded project
under Contract No. 20090460736.
B.~X.~ is supported by the
Director, Office of Energy Research, Office of High Energy and
Nuclear Physics, Divisions of Nuclear Physics, of the U.S.
Department of Energy under Contract No. DE-AC02-05CH11231.
\end{acknowledgments}

\end{document}